\documentclass[%
reprint,
superscriptaddress,
showpacs,
amsmath,amssymb,
prc,
floatfix, ]%
{revtex4-1}

\usepackage{color}

\usepackage{graphicx}
\usepackage{dcolumn}
\usepackage{bm}
\usepackage[dvipdfm,bookmarks=true,colorlinks,%
            citecolor=blue,linkcolor=blue,hypertex, %
            breaklinks=true]{hyperref}

\begin{document}

\hyphenation{Ref}


\title{Quadrupole deformation $(\beta,\gamma)$ of light $\Lambda$ hypernuclei 
       in constrained relativistic mean field model: shape evolution and 
       shape polarization effect of $\Lambda$ hyperon}

\author{Bing-Nan Lu}%
 \affiliation{Key Laboratory of Frontiers in Theoretical Physics,
              Institute of Theoretical Physics, Chinese Academy of Sciences,
              Beijing 100190, China}
\author{En-Guang Zhao}%
 \affiliation{Key Laboratory of Frontiers in Theoretical Physics,
              Institute of Theoretical Physics, Chinese Academy of Sciences,
              Beijing 100190, China}
 \affiliation{Center of Theoretical Nuclear Physics, National Laboratory
              of Heavy Ion Accelerator, Lanzhou 730000, China}
 \affiliation{School of Physics, Peking University,
              Beijing 100871, China}
\author{Shan-Gui Zhou}%
 \email{sgzhou@itp.ac.cn}
 \affiliation{Key Laboratory of Frontiers in Theoretical Physics,
              Institute of Theoretical Physics, Chinese Academy of Sciences,
              Beijing 100190, China}
 \affiliation{Center of Theoretical Nuclear Physics, National Laboratory
              of Heavy Ion Accelerator, Lanzhou 730000, China}

\date{\today}

\begin{abstract}
The shapes of light normal nuclei and $\Lambda$ hypernuclei are investigated in the $(\beta, \gamma)$ deformation plane by using a newly developed constrained relativistic mean field (RMF) model. As examples, the results of some C, Mg, and Si nuclei are presented and discussed in details. We found that for normal nuclei the present RMF calculations and previous Skyrme-Hartree-Fock models predict similar trends of the shape evolution with the neutron number increasing. But some quantitative aspects from these two approaches, such as the depth of the minimum and the softness in the $\gamma$ direction, differ a lot for several nuclei. For $\Lambda$ hypernuclei, in most cases, the addition of a $\Lambda$ hyperon alters slightly the location of the ground state minimum towards the direction of smaller $\beta$ and softer $\gamma$ in the potential energy surface $E \sim (\beta, \gamma)$. There are three exceptions, namely, $^{13}_\Lambda$C, $^{23}_\Lambda$C, and $^{31}_\Lambda$Si in which the polarization effect of the additional $\Lambda$ is so strong that the shapes of these three hypernuclei are drastically different from their corresponding core nuclei.
\end{abstract}


\maketitle


\section{Introduction~\label{sec:intro}}

Since the first observation of hypernuclei in 1953~\cite{Danysz1953_PhilMag744-348}, a lot of experimental efforts have focused on the study of the spectroscopy of hypernuclei, see, for examples, Refs.~\cite{Hashimoto2006_PPNP57-564,Nagae2010_PTPS185-299,Tang2011_DSJLAB} for recent reviews. Due to the additional strangeness degree of freedom, a hyperon is free from nucleon's Pauli exclusion principle. Thus it can move deep inside the nuclei and may serve as an impurity for probing many nuclear properties that are not accessible by normal methods. The study of hypernuclei can also provide detailed and accurate information about the hyperon-hyperon (YY) and hyperon-nucleon (YN) interactions~\cite{Hao1993_PRL71-1498,Ma1996_NPA608-305,Tzeng2002_PRC65-047303,Hiyama2010_PTPS185-1} which are important not only for the understanding of hyper nuclear structure but also for the study of hyper matter and neutron stars~\cite{Hofmann2001_PRC64-025804}. 

As an impurity in normal nuclei, a hyperon may induce many effects on the core nucleus, such as the shrinkage of the size~\cite{Motoba1983_PTP70-189,Hiyama1999_PRC59-2351,Tanida2001_PRL86-1982,Tan2001_CPL18-1030,Hiyama2010_PRL104-212502}, the change of the shape which will be discussed later, the modification of its cluster structure~\cite{Hiyama1996_PRC53-2075}, the occurrence of nucleon and hyperon skin or halo~\cite{Hiyama1996_PRC53-2075,Lu2002_CPL19-1775,Lu2003_EPJA17-19}, and the shift of neutron drip line to a neutron-rich side~\cite{Vretenar1998_PRC57-R1060,Lu2003_EPJA17-19,Zhou2008_PRC78-054306}. 

The shape describes in an intuitive way the spatial density distribution of a quantum many-body system. Most of known nuclei are non-spherical and many are well-deformed as manifested by regular rotational spectra~\cite{Bohr1969_Nucl_Structure}. The shape-driven effect of valence nucleon(s) has been extensively studied in nuclear high-spin states, see, for examples, Refs.~\cite{Paul1988_PRL61-42,Nazarewicz1990_NPA512-61,Ma1996_HEPNP20-865,Zhou2007_PRC75-034314}. A well known example is that a nucleon occupying a high-$j$ and low-$\Omega$ orbital would drive the nucleus to a more prolate shape. It is expected that the addition of a hyperon may also result in a shape polarization effect. Since the additional hyperon is not restricted by the nucleons' Pauli exclusion principle, it tends to occupy the lowest $s$ orbital, thus driving the core nucleus to be more spherical. If it occupies a $p$ orbital, a hyperon may enhance the nuclear deformation~\cite{Isaka2011_1104.3940}.

The self-consistent mean field models, either the non-relativistic~\cite{Rayet1976_AoP102-226,*Rayet1981_NPA367-381,Lanskoy1997_PRC55-2330,*Lanskoy1998_PRC58-3351,Cugnon2000_PRC62-064308,Vidana2001_PRC64-044301} or the relativistic ones~\cite{Brockmann1977_PLB69-167,Bouyssy1982_NPA381-445,Mares1989_ZPA333-209,*Mares1990_PLB249-181,Rufa1990_PRC42-2469,Cohen1991_PRC44-1181,Glendenning1991_PRL67-2414,*Glendenning1993_PRC48-889,Sugahara1994_PTP92-803,Mares1994_PRC49-2472,Marcos1998_PRC57-1178,Keil2002_PRC66-054307,Shen2006_PTP115-325,Zhong2006_PRC74-034321,Song2009_CPL26-122102}, have been extensively used to reproduce the available hypernuclear data and/or make various predictions for hypernuclei. Up to now most of these studies focus on spherical systems. The first self-consistent mean field study of hypernuclei with an axially deformed Skyrme Hartree-Fock (SHF) model was finished by Zhou et al.~\cite{Zhou2007_PRC76-034312}. It was found that \textit{the core nuclei and the corresponding hypernuclei have similar deformations with the same sign}~\cite{Zhou2007_PRC76-034312} which means that the shape polarization effect of the $\Lambda$ hyperon is quite small. A further study within the same framework but with a microscopic $\Lambda$N force gives similar conclusions~\cite{Zhou2009_SCG52-1548}. However, a relativistic mean field (RMF) study reveals that although in most cases the results are similar to the SHF calculations, there are indeed several exceptions, for examples, $^{13}_{\Lambda}$C and $^{29}_{\Lambda}$Si whose shapes change dramatically compared to their corresponding core nuclei~\cite{Win2008_PRC78-054311}. The different results between the SHF and RMF calculations are attributed to the different polarization effect of the additional $\Lambda$ in these two approaches~\cite{Schulze2010_PTP123-569}. In this sense, the experimental information related to shapes of hypernuclei is much desirable and would be used as a good benchmark for theories.  

The triaxiality is an important shape degree of freedom in many nuclei. In an atomic nucleus with a stably triaxial shape, the spontaneous broken chiral symmetry occurs~\cite{Frauendorf1997_NPA617-131,Frauendorf2001_RMP73-463,Meng2010_JPG37-064025} and the wobbling motion is also expected~\cite{Odegard2001_PRL86-5866,Chen2011_NSC2010}. The triaxial deformation plays important roles in $\gamma$-soft nuclei and in nuclei in the transitional mass regions~\cite{Guo2007_PRC76-034317,Guo2007_PRC76-065801,Li2009_PRC80-061301}. The $\gamma$ deformation also changes considerably the local minima and the saddle point in the potential energy surface of heavy nuclei~\cite{Pashkevich1969_NPA133-400,Abusara2010_PRC82-044303}. Recently Win et al. have studied hypernuclei by using a SHF model with the triaxial degree of freedom included~\cite{Win2011_PRC83-014301}. It is found that with an additional $\Lambda$ hyperon no significant change occurs for the nuclear shapes except that the potential energy surface becomes softer in the $\gamma$ direction. 


So far the relativistic description of hypernuclei is only restricted to spherical or axially deformed cases. It is the aim of this paper to investigate the triaxial deformations of the $\Lambda$ hypernuclei and the shape polarization effect of the $\Lambda$ hyperon in the RMF model, as a comparative study with the SHF calculations~\cite{Win2011_PRC83-014301}. 

The paper is organized as follows. In Section~\ref{sec:theory} we briefly introduce the RMF model used in the hypernuclear studies with emphasis on the extension of the RMF model to the inclusion of the hyperon and the triaxiality. In Section~\ref{sec:results}, we present the calculated results for C, Mg, and Si isotopes and the corresponding hypernuclei and discuss the shape polarization effect of the $\Lambda$ hyperon. Finally a summary is given in Section~\ref{sec:summary}. 

\section{Triaxially deformed relativistic mean field model for hypernuclei~\label{sec:theory}}

In the relativistic mean field (RMF) model, the hadrons interact with each other via the exchange of $\sigma$, $\omega$, $\rho$ mesons and the photon. For hypernuclei, the RMF Lagrangian density can be written as:
\begin{equation}
 \mathcal{L} = \mathcal{L}_{0} + \mathcal{L}_{\Lambda} ,
 \label{eq:L}
\end{equation}
where $\mathcal{L}_{0}$ is the standard RMF Lagrangian density describing the nucleons and the couplings between nucleons and mesons~\cite{Serot1986_ANP16-1,Reinhard1989_RPP52-439,Ring1996_PPNP37-193,Vretenar2005_PR409-101,Meng2006_PPNP57-470} and $\mathcal{L}_{\Lambda}$ is that for the hyperon:
\begin{eqnarray}
 \mathcal{L}_{\Lambda} 
 & = & \bar{\psi}_{\Lambda} \left( i \gamma^{\mu} \partial_{\mu} - m_{\Lambda} 
                                 - g_{\sigma\Lambda} \sigma 
                                 - g_{\omega\Lambda} \gamma^{\mu} \omega_{\mu}
                            \right)
             \psi_{\Lambda}
 \nonumber \\
 &   & \mbox{}
 + \frac{f_{\omega\Lambda\Lambda}}{4m_{\Lambda}} 
   \bar{\psi}_{\Lambda} \sigma^{\mu\nu} \Omega_{\mu\nu} \psi_{\Lambda}
 ,
 \label{eq:LL}
\end{eqnarray}
where $m_{\Lambda}$ is the mass of the $\Lambda$ hyperon, $g_{\sigma\Lambda}$ and $g_{\omega\Lambda}$ are the coupling constants of the $\Lambda$ hyperon with the scalar and vector meson fields, respectively. The last term represents the tensor coupling between the $\Lambda$ hyperon and the $\omega$ field~\cite{Jennings1990_PLB246-325}. $\Omega_{\mu\nu}$ is the field tensor of the $\omega$ field defined as $\Omega_{\mu\nu}=\partial_{\mu}\omega_{\nu}-\partial_{\nu}\omega_{\mu}$. Couplings to the $\rho$ meson and the photon vanishes for $\Lambda$ hyperons which are neutral and isoscalar. 

Under the mean field approximation, the single particle Dirac equation for $\Lambda$ hyperons reads:
\begin{equation}
  \left[ \vec{\alpha} \cdot \vec{p}
       +\beta \left( m_{\Lambda} + S_{\Lambda} \right)
       +V_{\Lambda}
       +T_{\Lambda}
  \right] \psi_{\Lambda i}
  =
  \epsilon_{i} \psi_{\Lambda i}
,
\end{equation}
with the scalar potential $S_{\Lambda}=g_{\sigma\Lambda}\sigma$, the vector potential $V_{\Lambda}=g_{\omega\Lambda}\omega$ and the tensor potential:
\begin{equation}
  T_{\Lambda}
  =  - \frac{ f_{\omega\Lambda\Lambda} }{ 2m_{\Lambda} }
    \beta \left( \vec{\alpha} \cdot \vec{p} \right) \omega
.
\end{equation}

The potential energy surface (PES) is obtained by the constrained self-consistent calculation,
\begin{equation}
 E^{\prime} = \langle \hat{H} \rangle
            + \sum_{n=1}^{N_{c}} \frac{1}{2} C_{n}
              \left( \langle \hat{Q}_{n} \rangle - \mu_{n}
              \right)^{2}
 ,
\end{equation} 
where $\hat{H}$ is the RMF Hamiltonian, $\hat{Q}_{n}$'s are the multipole operators to be constrained and $N_{c}$ is the dimension of the constraining space. The quadrupole deformation parameters $\beta$ and $\gamma$ are calculated from the multipole moments of the baryon density distributions: 
\begin{equation}
 \beta
  = \sqrt{ \frac{\pi}{5} }
           \frac{ \sqrt{ \langle \hat{Q}_{20} \rangle^{2}
                       + 3\langle \hat{Q}_{22} \rangle^{2}
                       }
                }{A \langle r^{2} \rangle}
 ,
\end{equation}
\begin{equation}
 \gamma
 = \arctan \frac{ \sqrt{3} \langle \hat{Q}_{22} \rangle }
                { \langle \hat{Q}_{20} \rangle }
 ,
\end{equation} 
where $\langle\hat{Q}_{20}\rangle$ and $\langle\hat{Q}_{22}\rangle$
are the quadrupole moments: 
\begin{eqnarray}
 \langle \hat{Q}_{20} \rangle
 & = &\int d\tau \
           \rho \left( \vec{r} \right)
           \left( 3z^{2}-r^{2} \right)
,
\nonumber \\
 \langle \hat{Q}_{22} \rangle
 & = &\int d\tau \
           \rho \left( \vec{r} \right)
           \left( x^{2}-y^{2} \right)
 .
\end{eqnarray} 

For normal nuclei, the triaxially deformed RMF model has been developed based on expanding the nucleon Dirac spinor in a three-dimensional harmonic oscillator (3DHO) basis~\cite{Hirata1996_NPA609-131,Meng2006_PRC73-037303}. As an alternative approach, in the present work, the RMF equations are solved in an axially deformed harmonic oscillator (ADHO) basis~\cite{Gambhir1990_AoP198-132,Gambhir1990_AoP198-132}. We have modified the \verb+DIZ+ (or \verb+RMFAXIAL+) code~\cite{Gambhir1990_AoP198-132,Ring1997_CPC105-77} in order to allow the triaxial deformation and to include the hyperon. The basis wave functions are solutions of a Schr\"odinger equation with an ADHO potential:
\begin{eqnarray}
 \left(   -\frac{\hbar^{2}}{2M} \nabla^{2}
          +\frac{1}{2}M \left(  \omega_{r}^{2}r^{2}
                               +\omega_{z}^{2}z^{2}
                        \right) 
 \right) | \alpha \rangle      
 & = & E_{\alpha} | \alpha \rangle
,
\end{eqnarray}
with $r=\sqrt{x^2+y^2}$ and 
\begin{equation}
 | \alpha \rangle  =  \phi_{n_{z}}(z) R_{n_{r}}^{m}(r) 
                      \frac{1}{\sqrt{2\pi}} \exp \left( im\theta\right ) 
                      \chi_{s}
,
\end{equation}
where $\alpha=(n_{z}, n_{r}, m, s)$ are the asymptotic quantum numbers and $\chi_{s}$ is for the spin. They are characterized by the basis deformation $\beta_{\text{B}}$. These basis states form an orthonormal complete set and can be used to expand any spinor wave functions irrespective with their symmetries.  

The projection of the total angular momentum on the symmetric $z$-axis $K$ is not conserved due to the breaking of the axial symmetry. The remaining symmetries are discrete ones such as interchanges of the three axes. Namely, the system is invariant under the point group $D_2$. 

To describe the potentials and densities we use a discretized three-dimensional mesh in the space. These mesh points are selected so that a Gaussian quadrature can be applied in the $r$ and $z$ directions, while an equally distributed mesh is used for the azimuthal angle. For convenience we use the Fourier expansion of the potentials $V$ and densities $\rho$:
\begin{widetext}
\begin{equation}
 f\left( z, r, \theta \right) = f_{0}(z,r) \frac{1}{\sqrt{2\pi}}
          + \sum_{n=1}^{\infty} f_{n}(z,r) \frac{1}{\sqrt{\pi}} \cos\left( 2n\theta \right) 
,
\end{equation}
with $f = V$ or $\rho$. In an axially symmetric case only the $n=0$ term survives, the present code then returns back to \verb+DIZ+. Most of the formulas for $f_{n}$ are formally the same as the corresponding ones for $f_{0}$, for example, the matrix element of the potential $V$ between two basis states is:
\begin{eqnarray}
 V_{\alpha \alpha^{\prime}} 
 & = & 
 \langle n_{z}, n_{r}, m, s | V | n_{z}^{\prime}, n_{r}^{\prime}, m^{\prime}, s^{\prime} \rangle  
 \nonumber \\
 & = & 
 \delta_{s, s^{\prime}} \frac{1}{2\sqrt{\pi}}
 \left[  \sqrt{2} \delta_{K, K^{\prime}} 
          R^{0;m,m^\prime}_{n_z,n_r;n^\prime_z,n^\prime_r} 
       + \sum_{n=1}^{\infty}  
          \left( \delta_{K^{\prime}-K+2n,0} + \delta_{K^{\prime}-K-2n,0} \right)
          R^{n;m,m^\prime}_{n_z,n_r;n^\prime_z,n^\prime_r} 
  \right]
,
\end{eqnarray}
where
\begin{eqnarray}
 R^{n;m,m^\prime}_{n_z,n_r;n^\prime_z,n^\prime_r}
 & \equiv &
 \int_{-\infty}^{\infty}dz \int_{0}^{\infty} r dr \ 
   \phi_{n_{z}}(z)          R_{n_{r}}^{m}(r) 
   V_{n}(z,r) 
   \phi_{n_{z}^{\prime}}(z) R_{n_{r}^{\prime}}^{m^{\prime}}(r)
 , \
 n = 0, 1, \cdots 
 .
\end{eqnarray}
\end{widetext}

The Klein-Gordon equations for mesons are also solved by the basis expansion method, while the Coulomb field is solved by the Green's function method. 

\begin{table}
\caption{The calculated binding energies of $^{26}$Si against $N_\mathrm{F}$, the number of major shells for the Fermion basis. The parameter set PK1 is used. The deformation is constrained to $\beta=0.4$ and $\gamma=$ 10$^{\circ}$, 30$^{\circ}$, and 50$^{\circ}$, respectively. The basis for bosons are truncated up to  $N_{B}=20$. The unit for energies is MeV.~\label{tab:convergence}}
\begin{centering}
\begin{tabular}{c>{\centering}p{0.2cm}rrr}
\hline 
\hline
$N_\mathrm{F}$\textbackslash{}
$(\beta,\gamma)$  &  & (0.4,10$^{\circ}$) & (0.4,30$^{\circ}$) & (0.4,50$^{\circ}$)\tabularnewline
\hline
 8 &  & $-$201.667 & $-$202.270 & $-$200.803\tabularnewline
10 &  & $-$201.523 & $-$202.093 & $-$200.633\tabularnewline
12 &  & $-$201.382 & $-$201.952 & $-$200.497\tabularnewline
14 &  & $-$201.325 & $-$201.907 & $-$200.468\tabularnewline
16 &  & $-$201.296 & $-$201.881 & $-$200.453\tabularnewline
18 &  & $-$201.294 & $-$201.878 & $-$200.453\tabularnewline
\hline
\hline
\end{tabular}
\par\end{centering}
\end{table}

In order to get a point on the PES with given deformation parameters $(\beta,\gamma)$, the axial deformation parameter of the basis is set to be, 
\begin{equation}
 \beta_{\text{B}} = \beta \cos \gamma
.
\end{equation}
The harmonic oscillator basis are truncated up to $N_\mathrm{F}$ fermion shells and $N_\mathrm{B}$ boson shells. The convergence of our method is checked for the nucleus $^{26}$Si. Table~\ref{tab:convergence} shows the calculated binding energies using different $N_\mathrm{F}$. Because the time consumption of the code is less affected by $N_\mathrm{B}$, we set $N_\mathrm{B}=20$ which is big enough. Three typical points on the PES are chosen, i.e., $(\beta,\gamma)=$ (0.4,10$^{\circ}$), (0.4,30$^{\circ}$), (0.4,50$^{\circ}$). In each column in Table~\ref{tab:convergence} the same calculation is performed with $N_\mathrm{F} = 8,\ 10,\ \cdots,\ 18$. The truncation errors are less than 100 keV for $N_\mathrm{F} \geq 12$ and less than 30 keV for $N_\mathrm{F} \geq 14$ for these three points. Furthermore, when discussing the energy differences on which we focus in the present work, the truncation errors may even be less due to the cancellation. Therefore we use $N_\mathrm{F}=14$ in the following calculations. 

\section{Results and discussions~\label{sec:results}}

\subsection{Numerical details}

\begin{table}
\caption{The RMF parameter sets used in the calculations.~\label{tab:forces}}
\begin{centering}
\begin{tabular}{llcccc}
\hline 
\hline
 & NN channel & $m_{\Lambda}$ (MeV) & $R_{\sigma}$ & $R_{\omega}$ & $R_{\omega\Lambda\Lambda}$
\tabularnewline
\hline
PK1-Y1~\protect\cite{Song2010_IJMPE19-2538} & PK1~\protect\cite{Long2004_PRC69-034319}  & 1115.6 & 0.580 & 0.620 & $-$1 \tabularnewline
NLSH-A~\protect\cite{Win2008_PRC78-054311} & NLSH~\protect\cite{Sharma1993_PLB312-377} & 1115.6 & 0.621 & 0.667 & $-$1 \tabularnewline
\hline
\hline
\end{tabular}
\end{centering}
\end{table}

In this work we adopted for the Lagrangian density (\ref{eq:L}) two parameter sets which are listed in Table~\ref{tab:forces}. For convenience we give in Table~\ref{tab:forces} the coupling constants by three dimensionless quantities defined as $R_{\sigma}=g_{\sigma\Lambda}/g_{\sigma}$, $R_{\omega}=g_{\omega\Lambda}/g_{\omega}$ and $R_{\omega\Lambda\Lambda}=f_{\omega\Lambda\Lambda}/g_{\omega\Lambda}$. The effective interaction PK1-Y1 is newly proposed by fitting to the experimental single-$\Lambda$ binding energies and $\Lambda$ spin-orbit splitting~\cite{Song2010_IJMPE19-2538}. Based on the parameter set PK1~\cite{Long2004_PRC69-034319} in the NN channel, PK1-Y1 can reproduce the binding energies of hypernuclei very well. For comparison we have also made calculations using the parameter set labeled as NLSH-A which is based on the NLSH parameter set for the nucleon-meson coupling constants~\cite{Sharma1993_PLB312-377} and has been used in the axially symmetric RMF calculations for the hypernuclei in Ref.~\cite{Win2008_PRC78-054311}.

We use a BCS scheme with constant gaps for the pairing. Following Ref.~\cite{Moeller1992_NPA536-20} the pairing gaps are taken as:
\begin{equation}
 \Delta_{\text{n}} = 4.8/N^{1/3}\textrm{ MeV}
,
 \quad
 \Delta_{\text{p}} = 4.8/Z^{1/3}\textrm{ MeV}
.
\end{equation}
In Ref.~\cite{Karatzikos2010_PLB689-72}, the fission barriers in actinides and superheavy nuclei are calculated by using different pairing schemes and it is found that BCS calculations with constant pairing gaps do not provide an adequate description of the fission barriers. The reason is if, e.g., a constant pairing strength $G$ is used, the resulting pairing gap changes considerably with deformation because the density of single particle levels around the Fermi surface does so. This is a very important conclusion, especially for the study of fission barriers in heavy nuclei. For the shape evolution and shape polarization effect of the $\Lambda$ hyperon in light nuclei we investigate here, the pairing does not play such a decisive role. In Ref.~\cite{Win2008_PRC78-054311} results from a constant pairing gap and a constant pairing strength $G$ are compared for the light hypernuclei in the RMF+BCS model. It is shown that most of the results are only slightly changed and the main features and conclusions remain the same. However, if the potential energy surface is rather soft, the minimum or minima in the potential energy surface would be different when different treatments of the pairing correlation are used. In such cases the configuration mixing effects must be also included, as discussed later.  

The center of mass correction is included either phenomenologically (for the parameter sets NLSH and NLSH-A) or microscopically (for the parameter sets PK1 and PK1-Y1). Note that neither the tensor force nor the center of mass correction has significant influence on shapes of the nuclei investigated in this work. For normal nuclei and single-$\Lambda$ hypernuclei, they only shift the PES's by roughly a few MeV as a whole. But when discussing the absolute value of the energies, their contributions are certainly not negligible. 

Using this triaxially deformed RMF+BCS method, we have calculated the PES's of even-even C, O, Ne, Mg, Si and S isotopes. Then by adding one $\Lambda$ hyperon, we have investigated in details the changes of the PES's of the corresponding hypernuclei. Next we take carbon and silicon isotopes as examples and examine the shape evolution in these two isotopic chains and the shape polarization effect of $\Lambda$ hyperon. Some results of $^{26}$Mg and $^{27}_\Lambda$Mg are also given for making a comparison with $^{26}$Si and $^{27}_\Lambda$Si. 

\subsection{Carbon hypernuclei~\label{sec:results-carbon}}

\begin{table*}
\caption{\label{tab:gs_C}The deformation parameters $\beta$ and $\gamma$, root mean square radii $r$ and binding energies $E$ of the carbon nuclei and hypernuclei calculated with parameter set PK1-Y1. The subscripts n, p, $\Lambda$ and tot represent the corresponding quantities for neutron, proton, $\Lambda$ hyperon, and the whole nucleus, respectively. The single-$\Lambda$ separation energies $B_{\Lambda\text{cal}}$ are also presented for the hypernuclei.}
\centering{}\begin{tabular}{c>{\centering}p{3mm}>{\centering}cccc>{\centering}p{3mm}cccc>{\centering}p{3mm}cccc>{\centering}p{3mm}rr}
\hline\hline 
 Nucleus &  & \multicolumn{4}{c}{$\beta$} && \multicolumn{4}{c}{$\gamma$ (deg)} && \multicolumn{4}{c}{r. m. s. radii (fm)} && \multicolumn{2}{c}{Energies (MeV)} \\
 \cline{1-1} \cline{3-6} \cline{8-11} \cline{13-16} \cline{18-19}  
         &  & $\beta_{\text{n}}$ & $\beta_{\text{p}}$ & $\beta_{\Lambda}$ & $\beta_{\text{tot}}$ && $\gamma_{\text{n}}$ & $\gamma_{\text{p}}$ & $\gamma_{\Lambda}$ & $\gamma_{\text{tot}}$ &  & $r_{\text{n}}$ & $r_{\text{p}}$ & $r_{\Lambda}$ & $r_{\text{tot}}$  &  & $-E_{\text{cal}}$  & $B_{\Lambda\text{cal}}$ \\
\hline 
          $^{10}$C && 0.340 & 0.255 &       & 0.284 && 0    & 0    &      & 0    && 2.53 & 2.87 &      & 2.74 &&  56.404 &        \\
$_{\Lambda}^{11}$C && 0.260 & 0.216 & 0.060 & 0.218 && 0    & 0    & 0    & 0    && 2.47 & 2.79 & 2.43 & 2.65 &&  65.652 &  9.250 \\
          $^{12}$C && 0.207 & 0.213 &       & 0.210 && 60.0 & 60.0 &      & 60.0 && 2.53 & 2.57 &      & 2.55 &&  87.764 &        \\
$_{\Lambda}^{13}$C &&     0 &     0 &     0 &     0 && 0    & 0    & 0    & 0    && 2.46 & 2.50 & 2.27 & 2.47 &&  99.752 & 11.990 \\
          $^{14}$C &&     0 &     0 &       &     0 && 0    & 0    &      & 0    && 2.74 & 2.52 &      & 2.65 && 104.943 &        \\
$_{\Lambda}^{15}$C &&     0 &     0 &     0 &     0 && 0    & 0    & 0    & 0    && 2.73 & 2.50 & 2.37 & 2.62 && 117.152 & 12.209 \\
          $^{16}$C && 0.225 & 0.179 &       & 0.211 && 0    & 0    &      & 0    && 2.90 & 2.45 &      & 2.74 && 110.398 &        \\
$_{\Lambda}^{17}$C && 0.202 & 0.158 & 0.068 & 0.183 && 0    & 0    & 0    & 0    && 2.88 & 2.43 & 2.40 & 2.70 && 123.422 & 13.024 \\
          $^{18}$C && 0.258 & 0.272 &       & 0.261 && 35.0 & 40.6 &      & 36.4 && 3.23 & 2.57 &      & 3.03 && 114.874 &        \\
$_{\Lambda}^{19}$C && 0.246 & 0.256 & 0.106 & 0.243 && 34.8 & 40.5 & 35.7 & 36.2 && 3.20 & 2.54 & 2.44 & 2.97 && 128.434 & 13.560 \\
          $^{20}$C && 0.275 & 0.308 &       & 0.282 && 60.0 & 60.0 &      & 60.0 && 3.47 & 2.63 &      & 3.24 && 119.601 &        \\
$_{\Lambda}^{21}$C && 0.259 & 0.291 & 0.119 & 0.261 && 60.0 & 60.0 & 60.0 & 60.0 && 3.43 & 2.60 & 2.47 & 3.17 && 133.591 & 13.990 \\
          $^{22}$C && 0.178 & 0.268 &       & 0.193 && 60.0 & 60.0 &      & 60.0 && 3.64 & 2.65 &      & 3.40 && 120.735 &        \\
$_{\Lambda}^{23}$C &&     0 &     0 &     0 &     0 && 0    & 0    & 0    & 0    && 3.55 & 2.58 & 2.44 & 3.28 && 135.340 & 14.605 \\
\hline\hline
\end{tabular}
\end{table*}

\begin{figure*}
\begin{centering}
\includegraphics[width=0.8\textwidth]{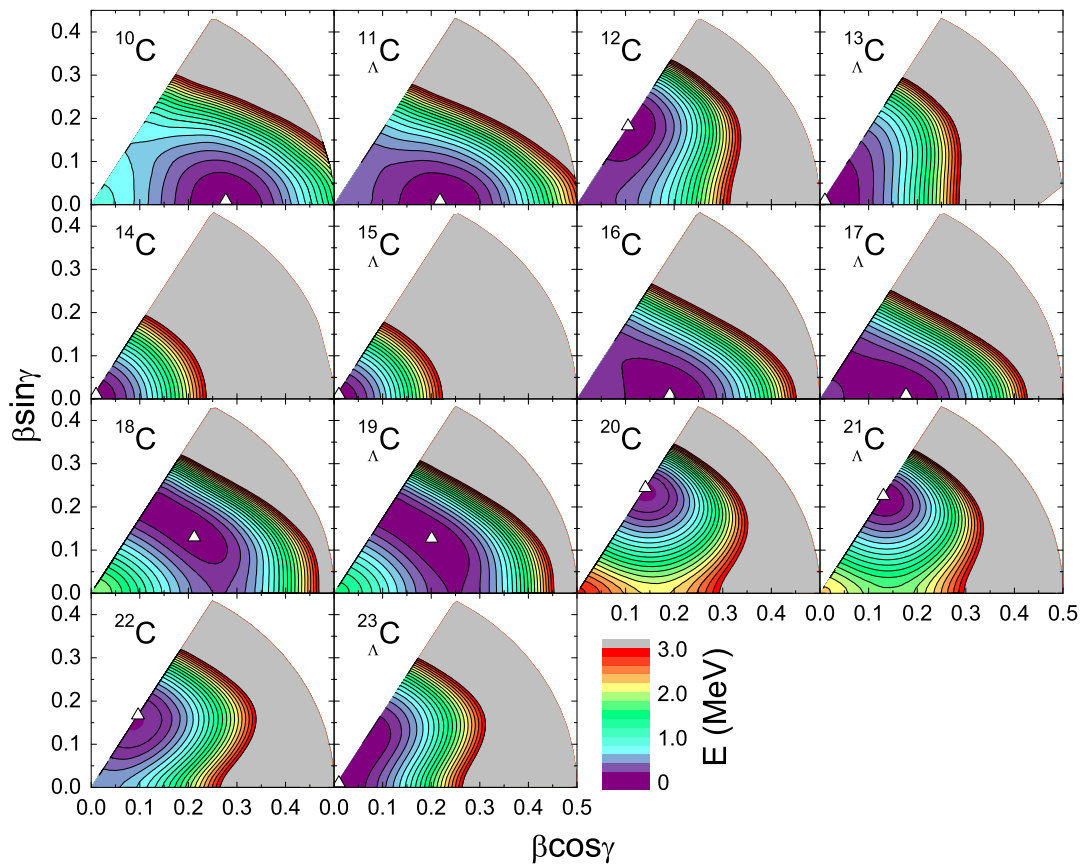} 
\par\end{centering}
\caption{\label{fig:C_forpaper}(Color online) The potential energy surfaces of carbon hypernuclei and the corresponding core nuclei in the $(\beta,\gamma)$ plane, calculated using the PK1-Y1 parameter set. The energies are normalized with respect to the binding energy of the absolute minimum. The contours join the points with the same energy. The contour interval is 0.15 MeV. The ground states are denoted by open triangles.}
\end{figure*}

The ground state properties of carbon isotopes have been studied extensively with the axially deformed RMF~\cite{Sharma1999_PRC59-1379,Lalazissis2004_EPJA22-37,Gangopadhyay2005_JPG31-1111} and SHF models~\cite{Sagawa2004_PRC70-054316} and the triaxially deformed SHF model~\cite{Zhang2008_PTP120-129}. Here we generalize the RMF calculations by considering two additional degrees of freedom: the triaxial degree of freedom for nuclear shape and the hyperon degree of freedom. We have performed constrained RMF calculations in the $(\beta,\gamma)$ plane for even-even carbon isotopes from $^{10}$C to the drip line nucleus $^{22}$C as well as for the hyper counterparts. Note that the study of carbon isotopes with triaxially deformed SHF models for normal nuclei~\cite{Zhang2008_PTP120-129} and hypernuclei~\cite{Win2011_PRC83-014301} are both available, while that with the RMF model is still absent.

The calculated ground state deformation parameters are summarized in Table~\ref{tab:gs_C}, together with the root mean square radii, binding energies and single $\Lambda$ separation energies. The single $\Lambda$ separation energy is defined as the energy difference between a hypernucleus and the corresponding core nucleus:
\begin{equation}
   B_{\Lambda} \left( _{\Lambda}^{A+1} \text{C} \right)
 = E \left( ^{A} \text{C} \right)
 - E \left( _{\Lambda}^{A+1} \text{C} \right)
.
\end{equation}
Firstly, the contraction due to the additional $\Lambda$ are observed for all the nuclei studies here, which is a manifestation of the glue like effect of the hyperon~\cite{Tanida2001_PRL86-1982}. Secondly, the radii of $^{10}$C and $_{\Lambda}^{11}$C are much larger than those of the neighbouring nuclei with two more neutrons. This is due to that they are close to the proton drip line. We mention that the earlier calculations with an axially deformed RMF+BCS model also show a similar trend~\cite{Sharma1999_PRC59-1379}. Finally, because the $\Lambda$ hyperon always occupies the lowest orbital generated by the mean field, the single $\Lambda$ separation energy $B_{\Lambda}$ can be seen as a measure of the depth and shape of the potential felt by the hyperon. In general this quantity should increase as the nucleon number increasing because the depth of the potential increases. This is clearly seen in Table~\ref{tab:gs_C}.

In Figure~\ref{fig:C_forpaper} we present the calculated PES's of carbon isotopes, together with those of the corresponding carbon hypernuclei with one additional $\Lambda$ hyperon. The locations of the ground states are denoted by open triangles. To unify the energy scales, we only show the relative energies with respect to the ground state. The contours join the points on the PES with the same energy. The energy difference between two neighbouring contours is 0.15 MeV. 

Many nuclei show two minima in both the prolate and the oblate sides of the PES's in axially deformed calculations. In the triaxially deformed calculations, at least one of them becomes a saddle point. In such cases, the relative energy differences among the minimum, the saddle point and the spherical configuration are key quantities characterizing the PES. We list in Table~\ref{tab:minC} the energies of the minimum or the saddle point in the prolate $E_\mathrm{prolate}$ or oblate sides $E_\mathrm{oblate}$ as well as that of the spherical configuration $E_\mathrm{spherical}$, with respect to the ground state energy. The results calculated with PK1-Y1 and NLSH-A parameter sets are both presented. 

It is convenient for further discussions to define the following two quantities. One is the deformation energy defined as the energy difference between the spherical shape and the ground state, i.e., $E_{\mathrm{def}} \equiv |E_{\mathrm{spherical}}-E_{\mathrm{ground}}|$. $E_{\mathrm{def}}$ characterizes the driving force to deformation qualitatively. The other is the prolate-oblate energy difference defined as the energy difference between the minimum in the prolate (oblate) side and the saddle point in the oblate (prolate) side for an axially deformed nucleus or the energy difference between two saddle points in the oblate and prolate sides for a triaxially deformed nucleus, i.e., $E_{\mathrm{po}} \equiv |E_{\mathrm{prolate}}-E_{\mathrm{oblate}}|$. Apparently it is only well defined if at least one saddle point exists in the prolate or oblate sides of the PES. $E_{\mathrm{po}}$ characterizes the softness of the PES in the $\gamma$ direction in most cases. That is, the larger $E_{\mathrm{po}}$ is, the steeper the PES in the $\gamma$ direction is.

\begin{table}
\caption{\label{tab:minC}The energies of the minimum or the saddle point in the prolate or oblate sides, and the energy of the spherical configuration, with respect to the ground state for carbon nuclei and hypernuclei, calculated with the PK1-Y1 and NLSH-A parameter sets. The energies are in MeV.}
\begin{centering}
\begin{tabular}{ccccccccc}
\hline 
\hline
 &  & \multicolumn{3}{c}{PK1-Y1} &  & \multicolumn{3}{c}{NLSH-A} \tabularnewline
 \cline{3-5} \cline{7-9} 
 &  & $E_{\text{prolate}}$ & $E_{\text{oblate}}$ & $E_{\text{sphercial}}$ 
 &  & $E_{\text{prolate}}$ & $E_{\text{oblate}}$ & $E_{\text{sphercial}}$ \tabularnewline
\hline
          $^{10}$C &  & 0    & 0.74 & 0.95 &  & 0    & 0.53 & 0.70 \tabularnewline
$_{\Lambda}^{11}$C &  & 0    & 0.34 & 0.37 &  & 0    & 0.25 & 0.29 \tabularnewline
          $^{12}$C &  & -    & 0    & 0.26 &  & -    & 0    & 0.11 \tabularnewline
$_{\Lambda}^{13}$C &  & -    & -    & 0    &  & -    & -    & 0    \tabularnewline
          $^{14}$C &  & -    & -    & 0    &  & -    & -    & 0    \tabularnewline
$_{\Lambda}^{15}$C &  & -    & -    & 0    &  & -    & -    & 0    \tabularnewline
          $^{16}$C &  & 0    & 0.18 & 0.28 &  & 0    & 0.20 & 0.34 \tabularnewline
$_{\Lambda}^{17}$C &  & 0    & 0.12 & 0.18 &  & 0    & 0.14 & 0.22 \tabularnewline
          $^{18}$C &  & 0.30 & 0.04 & 1.95 &  & 0.22 & 0.04 & 1.86 \tabularnewline
$_{\Lambda}^{19}$C &  & 0.14 & 0.05 & 1.57 &  & 0.08 & 0.14 & 1.52 \tabularnewline
          $^{20}$C &  & 2.28 & 0    & 2.88 &  & 2.15 & 0    & 2.75 \tabularnewline
$_{\Lambda}^{21}$C &  & 1.83 & 0    & 2.28 &  & 1.74 & 0    & 2.14 \tabularnewline
          $^{22}$C &  & -    & 0    & 0.54 &  & -    & 0    & 0.25 \tabularnewline
$_{\Lambda}^{23}$C &  & -    & -    & 0    &  & -    & -    & 0    \tabularnewline
\hline
\hline
\end{tabular}
\par\end{centering}
\end{table}

\subsubsection{Shape evolution of carbon isotopes}

First let us examine the shape evolution of the carbon isotopes. The isotopic dependence of the deformation with triaxiality has been investigated with the SHF model in Ref.~\cite{Zhang2008_PTP120-129}, where the PES's of even-even carbon nuclei in the $(\beta,\gamma)$ deformation plane are presented and discussed. Next one can find that there are some new features from the RMF calculations.

The energy minimum of $^{10}$C situated at $\beta\approx0.28$ and is rather stable against the triaxial distortion. The energy difference between the ground state and the saddle point on the oblate side $E_{\mathrm{po}}$ is 0.74 MeV, accounts for more than $1\%$ of the total binding energy. However, this quantity is less than 40 keV in the SHF calculations~\cite{Zhang2008_PTP120-129,Win2011_PRC83-014301}, indicating that $^{10}$C is rather $\gamma$-soft from the SHF calculations. To examine the parameter dependence of our results, we also performed the RMF calculation using the NLSH parameter set. The resulting $E_{\mathrm{po}}$ is 0.53 MeV. Results for $^{10}$C from other RMF parameter sets can also be found in Ref.~\cite{Sharma1999_PRC59-1379} where $E_{\mathrm{po}}$ is always larger than 0.5 MeV. Thus we conclude that the PES of $^{10}$C is apparently softer in SHF than in RMF in the $\gamma$ degree of freedom.

With two additional neutrons in the sd shell, $^{12}$C is driven to be oblate deformed. The driving force is so weak that the deformation energy $E_\mathrm{def}$ is only 0.26 MeV and 0.11 MeV from the PK1 and NLSH parameter sets, respectively. The SHF calculation gives a $E_\mathrm{def}$ as large as 1 MeV~\cite{Sagawa2004_PRC70-054316,Zhang2008_PTP120-129}. This nucleus is empirically known to be oblate in its ground state from the inelastic scattering experiments~\cite{Specht1971_NPA171-65,Yasue1983_NPA394-29,Simmonds1988_NPA482-653}, which is in consistent with our result. It is appropriate to compare the depth of the energy minimum obtained from different mean field models. Most of the recent SHF calculations of the carbon isotopes use a recipe that the spin-orbit interaction is reduced to 60\% of its original strength in order to reproduce the oblate shape of $^{12}$C. This prescription is rather arbitrary and it is known that the depth of the energy minimum with respect to the spherical configuration is sensitive to the spin-orbit interaction (see Fig.~1 in Ref.~\cite{Sagawa2004_PRC70-054316}). In Ref.~\cite{Schulze2010_PTP123-569} it is shown that using this reduction factor as an adjustable parameter, different shapes of a $\Lambda$ hypernucleus could be predicted. In contrast with the SHF method, the deformation of $^{12}$C is correctly reproduced without any adjustment of parameters in the RMF model here and in Ref.~\cite{Sharma1999_PRC59-1379}. It seems that the reduction of the spin-orbit interaction introduced in these SHF calculations is a little too strong, if the RMF results are reliable. This needs to be further explored.

With the neutron number $N=8$ which is magic, $^{14}$C is predicted to be spherical. When two and four more neutrons are added to the sd shell, $^{16}$C and  $^{18}$C turn to be prolate and triaxially deformed, respectively. For $^{16}$C, $E_{\mathrm{po}}$ is 0.18 MeV with the PK1 parameter set and 0.20 MeV with the NLSH parameter set while it is predicted to be 0.61 MeV in the SHF calculation~\cite{Sagawa2004_PRC70-054316}. Contrary to the result of $^{10}$C, the energy minimum of $^{16}$C is much deeper in SHF than in RMF. The softness of the PES of $^{16}$C is also obtained by using three different mean field models in Ref.~\cite{Buervenich2008_JPG35-025103}. $^{18}$C is the only nucleus with a triaxial deformation in carbon isotopes. There are saddle points at both the prolate and the oblate sides. The RMF calculation predicts that the oblate one is lower than the prolate one by 0.26 MeV, while from SHF calculations the prolate one is lower (cf. Fig.~\ref{fig:C_forpaper} here and Fig.~5 in Ref.~\cite{Zhang2008_PTP120-129}). The PES's of $^{16}$C and $^{18}$C are rather $\gamma$-soft. The ground state deformations of such nuclei with extremely soft PES's may not be well described in the mean field level because the ground state wave functions are always a strongly correlated superposition of different shapes with nearly the same energies. This suggests that further investigations of these nuclei should include beyond mean field effects by using the generator coordinate method~\cite{Bender2003_RMP75-121,Yao2010_PRC81-044311,*Yao2011_PRC83-014308}.

The shape evolves again to be oblate for $^{20}$C and $^{22}$C with the latter to be the last bound nucleus within the neutron drip line of carbon isotopes. $^{20}$C is strongly deformed and its PES is the steepest one among the carbon isotopes investigated here in either $\beta$ or $\gamma$ directions. $^{22}$C is suggested to be a halo nucleus according to the measured large enhancement of the reaction cross section for it compared to those for neighboring carbon isotopes~\cite{Tanaka2010_PRL104-062701}. We certainly can not reproduce the halo structure for $^{22}$C because in the present work a Harmonic Oscillator basis is used which is not able to give the large spatial density distributions in halo nuclei~\cite{Zhou2000_CPL17-717,Stoitsov2000_PRC61-034311,Zhou2003_PRC68-034323,Zhou2010_PRC82-011301R}.

The shape evolution of normal carbon nuclei can be roughly explained by examining the shell structure. One can see from the Nilsson diagram that 10 and 14 are prolate and oblate magic numbers respectively~\cite{Ring1980}, which are responsible for the prolate shape of $^{16}$C and oblate shape of $^{20}$C. 

In the above discussions we see that the RMF and SHF models~\cite{Zhang2008_PTP120-129} predict the same trend of the shape evolution with the neutron number increasing. But some quantitative aspects from these two approaches, such as the depth of the minimum and the softness in the $\gamma$ degree of freedom, differ a lot for some nuclei. 

\subsubsection{Shape polarization effect of $\Lambda$ hyperon in carbon hypernuclei}

Next let us discuss the shape evolution in carbon hypernuclei and the shape polarization effect of the $\Lambda$ hyperon. As is indicated in Ref.~\cite{Schulze2010_PTP123-569}, the influence of a hyperon on the PES can be as large as 1 MeV, the shape may evolve in a different way compared with normal ones. 

Roughly speaking, the deformations of the carbon hypernuclei are similar to their corresponding core nuclei, with two exceptions $_{\Lambda}^{13}$C and $_{\Lambda}^{23}$C. These two nuclei become spherical while the corresponding core nuclei $^{12}$C and $^{22}$C are both oblate. The spherical shape of $_{\Lambda}^{13}$C has been predicted in an axially deformed RMF calculation~\cite{Win2008_PRC78-054311}. In our triaxially deformed RMF model this is confirmed in the $(\beta,\gamma)$ deformation plane. As we discussed earlier, the deformation energy of $^{12}$C is quite small. From Fig.~\ref{fig:C_forpaper} one can see that at the oblate ($\gamma=60^{\circ}$) edge $^{12}$C is very soft in the $\beta$ direction. The additional $\Lambda$ drives $_{\Lambda}^{13}$C to be spherical. $_{\Lambda}^{23}$C becomes spherical but it is also very soft in the $\beta$ direction at the oblate edge. The calculation with NLSH-A parameter set also predicts spherical ground states for $_{\Lambda}^{13}$C and $_{\Lambda}^{23}$C. 

In the SHF calculations, the PES of $^{10}$C with a prolate shape is so soft in the $\gamma$ degree of freedom that one additional $\Lambda$ drives the shape of $_{\Lambda}^{11}$C to be oblate~\cite{Win2011_PRC83-014301}. In the present work, $_{\Lambda}^{11}$C is still prolate with a smaller $\beta$ compared to $^{10}$C. The addition of a $\Lambda$ hyperon only makes the PES of $_{\Lambda}^{11}$C a little softer than that of $^{10}$C. Similar situation holds for $_{\Lambda}^{17}$C. For the only triaxially deformed carbon nucleus, $^{18}$C, it is observed that the $\Lambda$ hyperon also makes the PES of $_{\Lambda}^{19}$C softer and the $\gamma$ softness increases more towards the prolate direction. Since $^{20}$C is strongly oblate deformed, the addition of a $\Lambda$ hyperon does not change its shape much. Interestingly, with one $\Lambda$ added, the PES of $_\Lambda^{15}$C becomes stiffer around the spherical minimum compared to its core nucleus $^{14}$C. This is also due to the spherical-driven effect of the $\Lambda$ hyperon.

The softness of the PES of a nucleus in the $\gamma$ direction can be measured by $E_{\mathrm{po}}$. For example, it is 0.74 MeV for $^{10}$C and 0.34 MeV for $_{\Lambda}^{11}$ C, which means that the PES of the latter is much softer than the former. However, this difference for $_{\Lambda}^{17}$C is only 0.06 MeV smaller than that for the corresponding core nucleus, which indicates a very tender change. 

From the above discussion it is seen that the shape evolution of carbon isotopes is modified due to the additional $\Lambda$ hyperon. On one hand, the transition point from the deformed shape to the spherical shape is shifted from $^{14}$C in normal nuclei to $^{13}_\Lambda$C in hypernuclei. On the other hand, an abrupt change from a strongly oblate shape to a spherical shape is observed for $_{\Lambda}^{23}$C. Thus the spontaneous symmetry breaking effect in carbon hypernuclei is very different from that in normal nuclei.

\subsection{Silicon hypernuclei~\label{sec:results-silicon}}

\begin{table}
\caption{\label{tab:minSi}The energies of the minimum or the saddle point in the prolate or oblate sides, and the energy of the spherical configuration, with respect to the ground state for silicon nuclei and hypernuclei, calculated with the PK1-Y1 and NLSH-A parameter sets. The unit of the energies is MeV. The results for $^{26}$Mg and $^{27}_{\Lambda}$Mg are also presented.}
\begin{tabular}{ccccccccc}
\hline \hline
 &  & \multicolumn{3}{c}{PK1-Y1} &  & \multicolumn{3}{c}{NLSH-A} \\
 \cline{3-5} \cline{7-9} 
 &  & $E_{\text{oblate}}$ & $E_{\text{prolate}}$ & $E_{\text{spherical}}$ 
 &  & $E_{\text{oblate}}$ & $E_{\text{prolate}}$ & $E_{\text{spherical}}$ \\
\hline
          $^{22}$Si &  & -       & -    & 0    &  & -    & -    & 0    \\
$_{\Lambda}^{23}$Si &  & -       & -    & 0    &  & -    & -    & 0    \\
          $^{24}$Si &  & 0.63    & 0    & 1.17 &  & 0.17 & 0    & 0.62 \\
$_{\Lambda}^{25}$Si &  & 0.40    & 0    & 0.77 &  & 0.07 & 0    & 0.33 \\
          $^{26}$Si &  & 1.02    & 0    & 2.20 &  & 0.45 & 0    & 1.32 \\
$_{\Lambda}^{27}$Si &  & 0.79    & 0    & 1.55 &  & 0.41 & 0    & 0.87 \\
          $^{28}$Si &  & 0       & -    & 1.04 &  & 0    & -    & 0.40 \\
$_{\Lambda}^{29}$Si &  & 0       & -    & 0.10 &  & -    & -    & 0    \\
          $^{30}$Si &  & 0       & 0.17 & 0.53 &  & 0    & 0.03 & 0.32 \\
$_{\Lambda}^{31}$Si &  & $<$0.01 & 0    & 0.12 &  & 0.05 & 0    & 0.14 \\
          $^{32}$Si &  & 0       & 0.61 & 0.66 &  & 0    & 0.49 & 0.52 \\
$_{\Lambda}^{33}$Si &  & 0       & 0.41 & 0.44 &  & 0    & 0.32 & 0.34 \\
          $^{34}$Si &  & -       & -    & 0    &  & -    & -    & 0    \\
$_{\Lambda}^{35}$Si &  & -       & -    & 0    &  & -    & -    & 0    \\
          $^{26}$Mg &  & 0.34    & 0.04 & 1.38 &  & 0.32 & 0.01 & 1.08 \\
$_{\Lambda}^{27}$Mg &  & 0.29    & 0    & 0.89 &  & 0.33 & 0    & 0.70 \\
\hline \hline
\end{tabular}
\end{table}
\begin{figure*}
\begin{centering}
\includegraphics[width=0.8\textwidth]{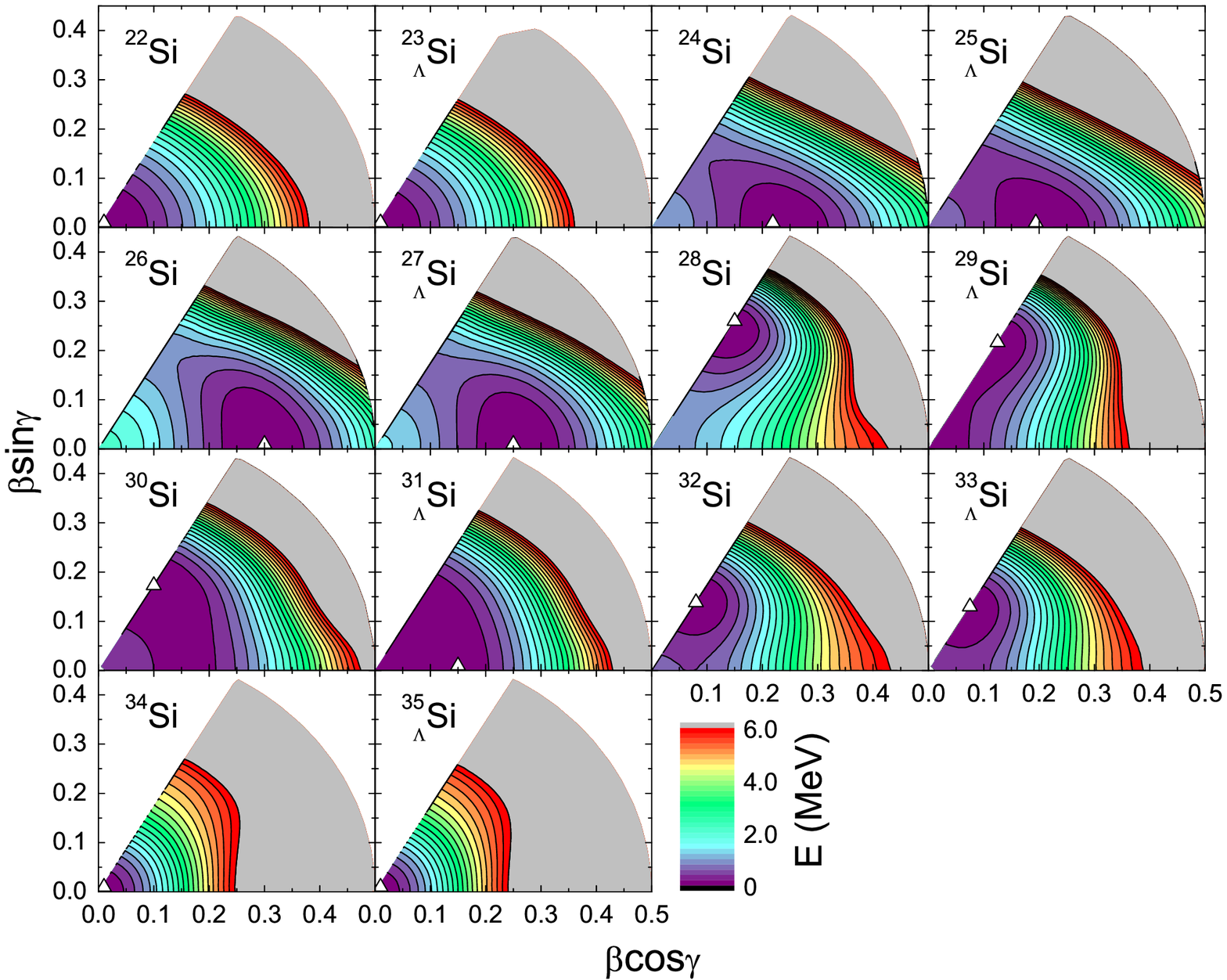}
\par\end{centering}
\caption{\label{fig:Si_forpaper}(Color online) The potential energy surfaces of silicon hypernuclei and the corresponding core nuclei in the $(\beta,\gamma)$ plane, calculated using the PK1-Y1 parameter set. The energies are normalized with respect to the binding energy of the absolute minimum. The contours join the points with the same energy. The contour interval is 0.3 MeV. The ground states are denoted by open triangles.}
\end{figure*}

\begin{figure}
\begin{centering}
\includegraphics[width=0.48\textwidth]{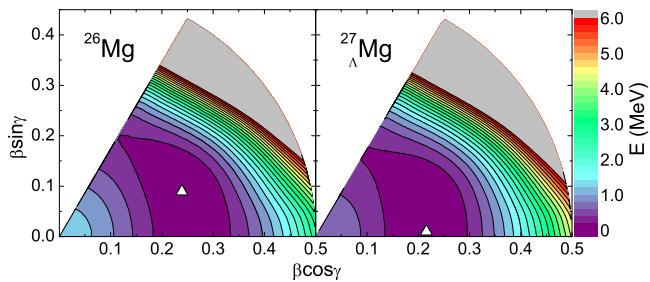}
\par\end{centering}
\caption{\label{fig:Mg26}(Color online) The potential energy surfaces of $^{26}$Mg and $_{\Lambda}^{27}$Mg in the $(\beta,\gamma)$ plane calculated using the PK1-Y1 parameter set. The energies are normalized with respect to the binding energy of the absolute minimum. The contours join the points with the same energy. The contour interval is 0.3 MeV. The ground states are denoted by open triangles.}
\end{figure}

As another example for the study of the shape evolution between two shell closures, we present in Fig.~\ref{fig:Si_forpaper} the PES's of silicon isotopes and the corresponding one-$\Lambda$ hypernuclei in the $(\beta,\gamma)$ plane. Because in silicon isotopes the proton number is 14 which energetically favors the oblate shape, the competition between the neutron and proton deformation driving forces may produce various types of PES's. When a $\Lambda$ hyperon is added, subtle changes are expected.

We list the energies of the minimum or the saddle point at the prolate and oblate sides and the spherical configuration, with respect to the ground state in Table~\ref{tab:minSi}. As the mirror nucleus of $^{26}$Si, the results for $^{26}$Mg are also presented. For silicon isotopes the parameter dependence of the results is a bit larger than that for carbon, but the results from these two different parameter sets are still in consistent with each other.

Starting from the spherical magic number nucleus $^{22}$Si, the shape evolves to be prolate in $^{24}$Si and $^{26}$Si. These two nuclei are rather $\gamma$ soft, similar as $^{16}$C and $^{18}$C. In the previous SHF calculations~\cite{Win2011_PRC83-014301} an oblate ground state was obtained for $^{26}$Si. Following Ref.~\cite{Win2011_PRC83-014301} we also calculate $^{26}$Mg which is the mirror nucleus of $^{26}$Si. The PES of $^{26}$Mg is presented in Fig.~\ref{fig:Mg26}. The ground state of $^{26}$Mg is triaxially deformed. Although the calculated PES's for $^{26}$Mg from the RMF and SHF calculations are both flat against the $\gamma$ deformation, our RMF model predicts a lower saddle point at the prolate side while the SHF method does at the oblate side (cf. Fig.~\ref{fig:Mg26} here and Fig.~12 in Ref.~\cite{Win2011_PRC83-014301}). For hypernuclei $^{25}_\Lambda$Si and $^{27}_\Lambda$Si, the values of the deformation parameter $\beta$ are a bit smaller than those of the corresponding core nuclei, respectively and the PES's become softer along the $\gamma$ direction. 

With two more neutrons in the sd shell, $^{28}$Si turns to be oblate with $E_{\mathrm{def}}$ as large as 1.04 MeV. Adding one $\Lambda$ hyperon results in a shape coexistence. Namely, the ground state is still oblate, but the energy of the spherical configuration with respect to the ground state is lowered to as small as 0.1 MeV. The barrier height between the two local minima is only about 0.25 MeV. The results discussed above are obtained from the parameter set PK1. Note that in the axially deformed RMF calculation in Ref.~\cite{Win2008_PRC78-054311}, $^{29}_{\Lambda}$Si is predicted to be spherical with the NLSH-A parameter set. When the NLSH-A parameter set is used in our calculation, we obtain the same conclusion as Ref.~\cite{Win2008_PRC78-054311}. As is seen in Table~\ref{tab:minSi}, for $^{28}$Si $E_{\mathrm{def}}$ is 0.40 MeV, but for $^{29}_{\Lambda}$Si $E_{\mathrm{def}}$ vanishes which means that $^{29}_{\Lambda}$Si is spherical. Therefore the prediction of the shape of $^{29}_{\Lambda}$Si is a bit parameter dependent. 

The PES of $^{30}$Si is almost the softest one among the nuclei investigated in the present work. Including an additional $\Lambda$ hyperon, the energy of $_{\Lambda}^{31}$Si is almost irrelevant with the deformation if $\beta<0.2$. The ground state moves from the oblate side to the prolate side. But the shift of the minimum does not mean much because the coherent superposition of the different shapes should be considered for the actual ground state. $^{32}$Si is also oblate and the PES is soft near the ground state. The $\Lambda$ hyperon softens slightly the PES of $_{\Lambda}^{33}$Si. Filling completely the neutron sd shell, stable spherical shapes are again obtained for $^{34}$Si and $^{35}_\Lambda$Si.

\section{Summary~\label{sec:summary}}

We developed a triaxially deformed RMF model for hypernuclei. Different from previous RMF calculations for normal nuclei, in the present work, the RMF equations are solved in an axially deformed harmonic oscillator (ADHO) basis. The convergence of the calculated results against the basis truncation is studied and it is shown that a reasonably large ADHO basis is able to provide desired accuracy in the triaxial RMF calculations. 

The shapes of C, O, Ne, Mg, Si and S $\Lambda$ hypernuclei are investigated in the $(\beta, \gamma)$ deformation plane by using this newly developed constrained RMF+BCS model with the parameter sets PK1-Y1 and NLSH-A. As examples, the results of some C, Mg, and Si $\Lambda$ hypernuclei are presented and we discussed in details the shape evolution of light normal nuclei and hypernuclei and the shape polarization effect of the $\Lambda$ hyperon. 

It is found that for normal nuclei the present RMF model and previous Skyrme-Hartree-Fock models predict similar trends of the shape evolution with the neutron number increasing. But some quantitative aspects from these two approaches, such as the depth of the minimum and $\gamma$ softness differ a lot for several nuclei. 

For $\Lambda$ hypernuclei, in most cases, the addition of a $\Lambda$ hyperon alters slightly the location of the ground state minimum towards the direction of smaller $\beta$ and softer $\gamma$ in the potential energy surface (PES) $E \sim (\beta, \gamma)$. There exist two exceptions in carbon isotopes, namely, $^{13}_\Lambda$C and $^{23}_\Lambda$C in which the polarization effect of the additional $\Lambda$ is so strong that it drives these nuclei from the oblate shape to spheres. Shape changes also occur in silicon isotopes. Although $^{28}$Si is oblate from both the PK1-Y1 and NLSH-A parameter sets, the prediction of the shape of $^{29}_{\Lambda}$Si is parameter dependent. $^{29}_{\Lambda}$Si is spherical with the NLSH-A parameter set but it is still oblate with the PK1-Y1 parameter set. Compared to the core nucleus, shape change also happens in $^{31}_\Lambda$Si, from an oblate shape in $^{30}$Si to a prolate one in $^{31}_\Lambda$Si. But the PES's of $^{30}$Si and $_{\Lambda}^{31}$Si are rather soft. The ground state deformation of a nucleus with such an extremely soft PES may not be well described in the mean field level because the ground state wave function should be a strongly correlated superposition of different shapes with nearly the same energies. This suggests that further investigations of these nuclei should include beyond mean field effects by using, for example, the generator coordinate method.

Finally we note that since different predictions about the shape polarization effect of the $\Lambda$ hyperon are made by different models and some times even by different effective interactions within the same model, the experimental information related to shapes of hypernuclei is highly desired and would be used as a good benchmark for theoretical models. In Ref.~\cite{Yao2011_1104.3200}, Yao et al. studied the impurity effect of the $\Lambda$ hyperon on collective excitations of nuclei based on potential energy surfaces calculated from the Skyrme HF model. It is found that the $\Lambda$ hyperon stretches the ground state band of the core of $_\Lambda^{25}$Mg and reduces the $B(E2:2^+_1 \rightarrow 0^+_1)$ value considerably due to a softening effect of $\Lambda$ on the potential energy surface. We would expect more profound effects in the collective spectrum from an additional $\Lambda$ if it changes even the shape of a nucleus, e.g., from a prolate shape in $^{12}$C to a spherical one in $_\Lambda^{13}$C. We expect that with the new or updated experimental facilities in J-PARC or JLab such measurements may become possible in the near future.

\begin{acknowledgments}
This work has been supported by NSFC (Grant Nos. 10875157, 10975100, and 10979066), MOST (973 Project 2007CB815000), and CAS (Grant Nos. KJCX2-EW-N01, KJCX2-SW-N17, and KJCX2-YW-N32). The computation of this work was supported by Supercomputing Center, CNIC of CAS. We thank S. N. Ershov, E. Hiyama, H. Lenske, V. V. Pashkevichfor, P. Ring, and J. M. Yao for helpful discussions and J. M. Yao for providing us the 3DHO RMF code to which we can compare our program. 
\end{acknowledgments}


\begin{thebibliography}{95}%
\makeatletter
\providecommand \@ifxundefined [1]{%
 \@ifx{#1\undefined}
}%
\providecommand \@ifnum [1]{%
 \ifnum #1\expandafter \@firstoftwo
 \else \expandafter \@secondoftwo
 \fi
}%
\providecommand \@ifx [1]{%
 \ifx #1\expandafter \@firstoftwo
 \else \expandafter \@secondoftwo
 \fi
}%
\providecommand \natexlab [1]{#1}%
\providecommand \enquote  [1]{``#1''}%
\providecommand \bibnamefont  [1]{#1}%
\providecommand \bibfnamefont [1]{#1}%
\providecommand \citenamefont [1]{#1}%
\providecommand \href@noop [0]{\@secondoftwo}%
\providecommand \href [0]{\begingroup \@sanitize@url \@href}%
\providecommand \@href[1]{\@@startlink{#1}\@@href}%
\providecommand \@@href[1]{\endgroup#1\@@endlink}%
\providecommand \@sanitize@url [0]{\catcode `\\12\catcode `\$12\catcode
  `\&12\catcode `\#12\catcode `\^12\catcode `\_12\catcode `\%12\relax}%
\providecommand \@@startlink[1]{}%
\providecommand \@@endlink[0]{}%
\providecommand \url  [0]{\begingroup\@sanitize@url \@url }%
\providecommand \@url [1]{\endgroup\@href {#1}{\urlprefix }}%
\providecommand \urlprefix  [0]{URL }%
\providecommand \Eprint [0]{\href }%
\providecommand \doibase [0]{http://dx.doi.org/}%
\providecommand \selectlanguage [0]{\@gobble}%
\providecommand \bibinfo  [0]{\@secondoftwo}%
\providecommand \bibfield  [0]{\@secondoftwo}%
\providecommand \translation [1]{[#1]}%
\providecommand \BibitemOpen [0]{}%
\providecommand \bibitemStop [0]{}%
\providecommand \bibitemNoStop [0]{.\EOS\space}%
\providecommand \EOS [0]{\spacefactor3000\relax}%
\providecommand \BibitemShut  [1]{\csname bibitem#1\endcsname}%
\let\auto@bib@innerbib\@empty
\bibitem [{\citenamefont {Danysz}\ and\ \citenamefont
  {Pniewski}(1953)}]{Danysz1953_PhilMag744-348}%
  \BibitemOpen
  \bibfield  {author} {\bibinfo {author} {\bibfnamefont {M.}~\bibnamefont
  {Danysz}}\ and\ \bibinfo {author} {\bibfnamefont {J.}~\bibnamefont
  {Pniewski}},\ }\href {\doibase 10.1080/14786440308520318} {\bibfield
  {journal} {\bibinfo  {journal} {Philos. Mag.}\ }\textbf {\bibinfo {volume}
  {44}},\ \bibinfo {pages} {348} (\bibinfo {year} {1953})}\BibitemShut
  {NoStop}%
\bibitem [{\citenamefont {Hashimoto}\ and\ \citenamefont
  {Tamura}(2006)}]{Hashimoto2006_PPNP57-564}%
  \BibitemOpen
  \bibfield  {author} {\bibinfo {author} {\bibfnamefont {O.}~\bibnamefont
  {Hashimoto}}\ and\ \bibinfo {author} {\bibfnamefont {H.}~\bibnamefont
  {Tamura}},\ }\href {\doibase 10.1016/j.ppnp.2005.07.001} {\bibfield
  {journal} {\bibinfo  {journal} {Prog. Part. Nucl. Phys.}\ }\textbf {\bibinfo
  {volume} {57}},\ \bibinfo {pages} {564} (\bibinfo {year} {2006})}\BibitemShut
  {NoStop}%
\bibitem [{\citenamefont {Nagae}(2010)}]{Nagae2010_PTPS185-299}%
  \BibitemOpen
  \bibfield  {author} {\bibinfo {author} {\bibfnamefont {T.}~\bibnamefont
  {Nagae}},\ }\href {\doibase 10.1143/PTPS.185.299} {\bibfield  {journal}
  {\bibinfo  {journal} {Prog. Theor. Phys. Suppl.}\ }\textbf {\bibinfo {volume}
  {185}},\ \bibinfo {pages} {299} (\bibinfo {year} {2010})}\BibitemShut
  {NoStop}%
\bibitem [{\citenamefont {Tang}\ \emph {et~al.}(2011)\citenamefont {Tang},
  \citenamefont {LeRose}, \citenamefont {Hashimoto}, \citenamefont {Nakamura},
  \citenamefont {Garibaldi}, \citenamefont {Markowitz},\ and\ \citenamefont
  {Reinhold}}]{Tang2011_DSJLAB}%
  \BibitemOpen
  \bibfield  {author} {\bibinfo {author} {\bibfnamefont {L.}~\bibnamefont
  {Tang}}, \bibinfo {author} {\bibfnamefont {J.~J.}\ \bibnamefont {LeRose}},
  \bibinfo {author} {\bibfnamefont {O.}~\bibnamefont {Hashimoto}}, \bibinfo
  {author} {\bibfnamefont {S.~N.}\ \bibnamefont {Nakamura}}, \bibinfo {author}
  {\bibfnamefont {F.}~\bibnamefont {Garibaldi}}, \bibinfo {author}
  {\bibfnamefont {P.}~\bibnamefont {Markowitz}}, \ and\ \bibinfo {author}
  {\bibfnamefont {J.}~\bibnamefont {Reinhold}},\ }in\ \href@noop {} {\emph
  {\bibinfo {booktitle} {Decade of Science at JLAB}}}\ (\bibinfo {year}
  {2011})\BibitemShut {NoStop}%
\bibitem [{\citenamefont {Hao}\ \emph {et~al.}(1993)\citenamefont {Hao},
  \citenamefont {Kuo}, \citenamefont {Reuber}, \citenamefont {Holinde},
  \citenamefont {Speth},\ and\ \citenamefont {Millener}}]{Hao1993_PRL71-1498}%
  \BibitemOpen
  \bibfield  {author} {\bibinfo {author} {\bibfnamefont {J.}~\bibnamefont
  {Hao}}, \bibinfo {author} {\bibfnamefont {T.~T.~S.}\ \bibnamefont {Kuo}},
  \bibinfo {author} {\bibfnamefont {A.}~\bibnamefont {Reuber}}, \bibinfo
  {author} {\bibfnamefont {K.}~\bibnamefont {Holinde}}, \bibinfo {author}
  {\bibfnamefont {J.}~\bibnamefont {Speth}}, \ and\ \bibinfo {author}
  {\bibfnamefont {D.~J.}\ \bibnamefont {Millener}},\ }\href {\doibase
  10.1103/PhysRevLett.71.1498} {\bibfield  {journal} {\bibinfo  {journal}
  {Phys. Rev. Lett.}\ }\textbf {\bibinfo {volume} {71}},\ \bibinfo {pages}
  {1498} (\bibinfo {year} {1993})}\BibitemShut {NoStop}%
\bibitem [{\citenamefont {Ma}\ \emph {et~al.}(1996{\natexlab{a}})\citenamefont
  {Ma}, \citenamefont {Speth}, \citenamefont {Krewald}, \citenamefont {Chen},\
  and\ \citenamefont {Reuber}}]{Ma1996_NPA608-305}%
  \BibitemOpen
  \bibfield  {author} {\bibinfo {author} {\bibfnamefont {Z.-Y.}\ \bibnamefont
  {Ma}}, \bibinfo {author} {\bibfnamefont {J.}~\bibnamefont {Speth}}, \bibinfo
  {author} {\bibfnamefont {S.}~\bibnamefont {Krewald}}, \bibinfo {author}
  {\bibfnamefont {B.-Q.}\ \bibnamefont {Chen}}, \ and\ \bibinfo {author}
  {\bibfnamefont {A.}~\bibnamefont {Reuber}},\ }\href {\doibase
  10.1016/0375-9474(96)00169-8} {\bibfield  {journal} {\bibinfo  {journal}
  {Nucl. Phys. A}\ }\textbf {\bibinfo {volume} {608}},\ \bibinfo {pages} {305}
  (\bibinfo {year} {1996}{\natexlab{a}})}\BibitemShut {NoStop}%
\bibitem [{\citenamefont {Tzeng}\ \emph {et~al.}(2002)\citenamefont {Tzeng},
  \citenamefont {Tzeng},\ and\ \citenamefont {Kuo}}]{Tzeng2002_PRC65-047303}%
  \BibitemOpen
  \bibfield  {author} {\bibinfo {author} {\bibfnamefont {Y.}~\bibnamefont
  {Tzeng}}, \bibinfo {author} {\bibfnamefont {S.~Y.~T.}\ \bibnamefont {Tzeng}},
  \ and\ \bibinfo {author} {\bibfnamefont {T.~T.~S.}\ \bibnamefont {Kuo}},\
  }\href {\doibase 10.1103/PhysRevC.65.047303} {\bibfield  {journal} {\bibinfo
  {journal} {Phys. Rev. C}\ }\textbf {\bibinfo {volume} {65}},\ \bibinfo
  {pages} {047303} (\bibinfo {year} {2002})}\BibitemShut {NoStop}%
\bibitem [{\citenamefont {Hiyama}\ \emph
  {et~al.}(2010{\natexlab{a}})\citenamefont {Hiyama}, \citenamefont {Motoba},
  \citenamefont {Rijken},\ and\ \citenamefont
  {Yamamoto}}]{Hiyama2010_PTPS185-1}%
  \BibitemOpen
  \bibfield  {author} {\bibinfo {author} {\bibfnamefont {E.}~\bibnamefont
  {Hiyama}}, \bibinfo {author} {\bibfnamefont {T.}~\bibnamefont {Motoba}},
  \bibinfo {author} {\bibfnamefont {T.~A.}\ \bibnamefont {Rijken}}, \ and\
  \bibinfo {author} {\bibfnamefont {Y.}~\bibnamefont {Yamamoto}},\ }\href
  {\doibase 10.1143/PTPS.185.1} {\bibfield  {journal} {\bibinfo  {journal}
  {Prog. Theor. Phys. Suppl.}\ }\textbf {\bibinfo {volume} {185}},\ \bibinfo
  {pages} {1} (\bibinfo {year} {2010}{\natexlab{a}})}\BibitemShut {NoStop}%
\bibitem [{\citenamefont {Hofmann}\ \emph {et~al.}(2001)\citenamefont
  {Hofmann}, \citenamefont {Keil},\ and\ \citenamefont
  {Lenske}}]{Hofmann2001_PRC64-025804}%
  \BibitemOpen
  \bibfield  {author} {\bibinfo {author} {\bibfnamefont {F.}~\bibnamefont
  {Hofmann}}, \bibinfo {author} {\bibfnamefont {C.~M.}\ \bibnamefont {Keil}}, \
  and\ \bibinfo {author} {\bibfnamefont {H.}~\bibnamefont {Lenske}},\ }\href
  {\doibase 10.1103/PhysRevC.64.025804} {\bibfield  {journal} {\bibinfo
  {journal} {Phys. Rev. C}\ }\textbf {\bibinfo {volume} {64}},\ \bibinfo
  {pages} {025804} (\bibinfo {year} {2001})}\BibitemShut {NoStop}%
\bibitem [{\citenamefont {Motoba}\ \emph {et~al.}(1983)\citenamefont {Motoba},
  \citenamefont {Bando},\ and\ \citenamefont {Ikeda}}]{Motoba1983_PTP70-189}%
  \BibitemOpen
  \bibfield  {author} {\bibinfo {author} {\bibfnamefont {T.}~\bibnamefont
  {Motoba}}, \bibinfo {author} {\bibfnamefont {H.}~\bibnamefont {Bando}}, \
  and\ \bibinfo {author} {\bibfnamefont {K.}~\bibnamefont {Ikeda}},\ }\href
  {\doibase 10.1143/PTP.70.189} {\bibfield  {journal} {\bibinfo  {journal}
  {Prog. Theor. Phys.}\ }\textbf {\bibinfo {volume} {70}},\ \bibinfo {pages}
  {189} (\bibinfo {year} {1983})}\BibitemShut {NoStop}%
\bibitem [{\citenamefont {Hiyama}\ \emph {et~al.}(1999)\citenamefont {Hiyama},
  \citenamefont {Kamimura}, \citenamefont {Miyazaki},\ and\ \citenamefont
  {Motoba}}]{Hiyama1999_PRC59-2351}%
  \BibitemOpen
  \bibfield  {author} {\bibinfo {author} {\bibfnamefont {E.}~\bibnamefont
  {Hiyama}}, \bibinfo {author} {\bibfnamefont {M.}~\bibnamefont {Kamimura}},
  \bibinfo {author} {\bibfnamefont {K.}~\bibnamefont {Miyazaki}}, \ and\
  \bibinfo {author} {\bibfnamefont {T.}~\bibnamefont {Motoba}},\ }\href
  {\doibase 10.1103/PhysRevC.59.2351} {\bibfield  {journal} {\bibinfo
  {journal} {Phys. Rev. C}\ }\textbf {\bibinfo {volume} {59}},\ \bibinfo
  {pages} {2351} (\bibinfo {year} {1999})}\BibitemShut {NoStop}%
\bibitem [{\citenamefont {Tanida}\ \emph {et~al.}(2001)\citenamefont {Tanida},
  \citenamefont {Tamura}, \citenamefont {Abe}, \citenamefont {Akikawa},
  \citenamefont {Araki}, \citenamefont {Bhang}, \citenamefont {Endo},
  \citenamefont {Fujii}, \citenamefont {Fukuda}, \citenamefont {Hashimoto},
  \citenamefont {Imai}, \citenamefont {Hotchi}, \citenamefont {Kakiguchi},
  \citenamefont {Kim}, \citenamefont {Kim}, \citenamefont {Miyoshi},
  \citenamefont {Murakami}, \citenamefont {Nagae}, \citenamefont {Noumi},
  \citenamefont {Outa}, \citenamefont {Ozawa}, \citenamefont {Saito},
  \citenamefont {Sasao}, \citenamefont {Sato}, \citenamefont {Satoh},
  \citenamefont {Sawafta}, \citenamefont {Sekimoto}, \citenamefont {Takahashi},
  \citenamefont {Tang}, \citenamefont {Xia}, \citenamefont {Zhou},\ and\
  \citenamefont {Zhu}}]{Tanida2001_PRL86-1982}%
  \BibitemOpen
  \bibfield  {author} {\bibinfo {author} {\bibfnamefont {K.}~\bibnamefont
  {Tanida}}, \bibinfo {author} {\bibfnamefont {H.}~\bibnamefont {Tamura}},
  \bibinfo {author} {\bibfnamefont {D.}~\bibnamefont {Abe}}, \bibinfo {author}
  {\bibfnamefont {H.}~\bibnamefont {Akikawa}}, \bibinfo {author} {\bibfnamefont
  {K.}~\bibnamefont {Araki}}, \bibinfo {author} {\bibfnamefont
  {H.}~\bibnamefont {Bhang}}, \bibinfo {author} {\bibfnamefont
  {T.}~\bibnamefont {Endo}}, \bibinfo {author} {\bibfnamefont {Y.}~\bibnamefont
  {Fujii}}, \bibinfo {author} {\bibfnamefont {T.}~\bibnamefont {Fukuda}},
  \bibinfo {author} {\bibfnamefont {O.}~\bibnamefont {Hashimoto}}, \bibinfo
  {author} {\bibfnamefont {K.}~\bibnamefont {Imai}}, \bibinfo {author}
  {\bibfnamefont {H.}~\bibnamefont {Hotchi}}, \bibinfo {author} {\bibfnamefont
  {Y.}~\bibnamefont {Kakiguchi}}, \bibinfo {author} {\bibfnamefont {J.~H.}\
  \bibnamefont {Kim}}, \bibinfo {author} {\bibfnamefont {Y.~D.}\ \bibnamefont
  {Kim}}, \bibinfo {author} {\bibfnamefont {T.}~\bibnamefont {Miyoshi}},
  \bibinfo {author} {\bibfnamefont {T.}~\bibnamefont {Murakami}}, \bibinfo
  {author} {\bibfnamefont {T.}~\bibnamefont {Nagae}}, \bibinfo {author}
  {\bibfnamefont {H.}~\bibnamefont {Noumi}}, \bibinfo {author} {\bibfnamefont
  {H.}~\bibnamefont {Outa}}, \bibinfo {author} {\bibfnamefont {K.}~\bibnamefont
  {Ozawa}}, \bibinfo {author} {\bibfnamefont {T.}~\bibnamefont {Saito}},
  \bibinfo {author} {\bibfnamefont {J.}~\bibnamefont {Sasao}}, \bibinfo
  {author} {\bibfnamefont {Y.}~\bibnamefont {Sato}}, \bibinfo {author}
  {\bibfnamefont {S.}~\bibnamefont {Satoh}}, \bibinfo {author} {\bibfnamefont
  {R.~I.}\ \bibnamefont {Sawafta}}, \bibinfo {author} {\bibfnamefont
  {M.}~\bibnamefont {Sekimoto}}, \bibinfo {author} {\bibfnamefont
  {T.}~\bibnamefont {Takahashi}}, \bibinfo {author} {\bibfnamefont
  {L.}~\bibnamefont {Tang}}, \bibinfo {author} {\bibfnamefont {H.~H.}\
  \bibnamefont {Xia}}, \bibinfo {author} {\bibfnamefont {S.~H.}\ \bibnamefont
  {Zhou}}, \ and\ \bibinfo {author} {\bibfnamefont {L.~H.}\ \bibnamefont
  {Zhu}},\ }\href {\doibase 10.1103/PhysRevLett.86.1982} {\bibfield  {journal}
  {\bibinfo  {journal} {Phys. Rev. Lett.}\ }\textbf {\bibinfo {volume} {86}},\
  \bibinfo {pages} {1982} (\bibinfo {year} {2001})}\BibitemShut {NoStop}%
\bibitem [{\citenamefont {Tan}\ \emph {et~al.}(2001)\citenamefont {Tan},
  \citenamefont {Luo}, \citenamefont {Ning},\ and\ \citenamefont
  {Ma}}]{Tan2001_CPL18-1030}%
  \BibitemOpen
  \bibfield  {author} {\bibinfo {author} {\bibfnamefont {Y.-H.}\ \bibnamefont
  {Tan}}, \bibinfo {author} {\bibfnamefont {Y.-A.}\ \bibnamefont {Luo}},
  \bibinfo {author} {\bibfnamefont {P.-Z.}\ \bibnamefont {Ning}}, \ and\
  \bibinfo {author} {\bibfnamefont {Z.-Y.}\ \bibnamefont {Ma}},\ }\href
  {\doibase 10.1088/0256-307X/18/8/311} {\bibfield  {journal} {\bibinfo
  {journal} {Chin. Phys. Lett.}\ }\textbf {\bibinfo {volume} {18}},\ \bibinfo
  {pages} {1030} (\bibinfo {year} {2001})}\BibitemShut {NoStop}%
\bibitem [{\citenamefont {Hiyama}\ \emph
  {et~al.}(2010{\natexlab{b}})\citenamefont {Hiyama}, \citenamefont {Kamimura},
  \citenamefont {Yamamoto},\ and\ \citenamefont
  {Motoba}}]{Hiyama2010_PRL104-212502}%
  \BibitemOpen
  \bibfield  {author} {\bibinfo {author} {\bibfnamefont {E.}~\bibnamefont
  {Hiyama}}, \bibinfo {author} {\bibfnamefont {M.}~\bibnamefont {Kamimura}},
  \bibinfo {author} {\bibfnamefont {Y.}~\bibnamefont {Yamamoto}}, \ and\
  \bibinfo {author} {\bibfnamefont {T.}~\bibnamefont {Motoba}},\ }\href
  {\doibase 10.1103/PhysRevLett.104.212502} {\bibfield  {journal} {\bibinfo
  {journal} {Phys. Rev. Lett.}\ }\textbf {\bibinfo {volume} {104}},\ \bibinfo
  {pages} {212502} (\bibinfo {year} {2010}{\natexlab{b}})}\BibitemShut
  {NoStop}%
\bibitem [{\citenamefont {Hiyama}\ \emph {et~al.}(1996)\citenamefont {Hiyama},
  \citenamefont {Kamimura}, \citenamefont {Motoba}, \citenamefont {Yamada},\
  and\ \citenamefont {Yamamoto}}]{Hiyama1996_PRC53-2075}%
  \BibitemOpen
  \bibfield  {author} {\bibinfo {author} {\bibfnamefont {E.}~\bibnamefont
  {Hiyama}}, \bibinfo {author} {\bibfnamefont {M.}~\bibnamefont {Kamimura}},
  \bibinfo {author} {\bibfnamefont {T.}~\bibnamefont {Motoba}}, \bibinfo
  {author} {\bibfnamefont {T.}~\bibnamefont {Yamada}}, \ and\ \bibinfo {author}
  {\bibfnamefont {Y.}~\bibnamefont {Yamamoto}},\ }\href {\doibase
  10.1103/PhysRevC.53.2075} {\bibfield  {journal} {\bibinfo  {journal} {Phys.
  Rev. C}\ }\textbf {\bibinfo {volume} {53}},\ \bibinfo {pages} {2075}
  (\bibinfo {year} {1996})}\BibitemShut {NoStop}%
\bibitem [{\citenamefont {L\"u}\ and\ \citenamefont
  {Meng}(2002)}]{Lu2002_CPL19-1775}%
  \BibitemOpen
  \bibfield  {author} {\bibinfo {author} {\bibfnamefont {H.-F.}\ \bibnamefont
  {L\"u}}\ and\ \bibinfo {author} {\bibfnamefont {J.}~\bibnamefont {Meng}},\
  }\href {\doibase 10.1088/0256-307X/19/12/310} {\bibfield  {journal} {\bibinfo
   {journal} {Chin. Phys. Lett.}\ }\textbf {\bibinfo {volume} {19}},\ \bibinfo
  {pages} {1775} (\bibinfo {year} {2002})}\BibitemShut {NoStop}%
\bibitem [{\citenamefont {L\"{u}}\ \emph {et~al.}(2003)\citenamefont {L\"{u}},
  \citenamefont {Meng}, \citenamefont {Zhang},\ and\ \citenamefont
  {Zhou}}]{Lu2003_EPJA17-19}%
  \BibitemOpen
  \bibfield  {author} {\bibinfo {author} {\bibfnamefont {H.~F.}\ \bibnamefont
  {L\"{u}}}, \bibinfo {author} {\bibfnamefont {J.}~\bibnamefont {Meng}},
  \bibinfo {author} {\bibfnamefont {S.~Q.}\ \bibnamefont {Zhang}}, \ and\
  \bibinfo {author} {\bibfnamefont {S.~G.}\ \bibnamefont {Zhou}},\ }\href
  {\doibase 10.1140/epja/i2002-10136-3} {\bibfield  {journal} {\bibinfo
  {journal} {Eur. Phys. J. A}\ }\textbf {\bibinfo {volume} {17}},\ \bibinfo
  {pages} {19} (\bibinfo {year} {2003})}\BibitemShut {NoStop}%
\bibitem [{\citenamefont {Vretenar}\ \emph {et~al.}(1998)\citenamefont
  {Vretenar}, \citenamefont {Poschl}, \citenamefont {Lalazissis},\ and\
  \citenamefont {Ring}}]{Vretenar1998_PRC57-R1060}%
  \BibitemOpen
  \bibfield  {author} {\bibinfo {author} {\bibfnamefont {D.}~\bibnamefont
  {Vretenar}}, \bibinfo {author} {\bibfnamefont {W.}~\bibnamefont {Poschl}},
  \bibinfo {author} {\bibfnamefont {G.~A.}\ \bibnamefont {Lalazissis}}, \ and\
  \bibinfo {author} {\bibfnamefont {P.}~\bibnamefont {Ring}},\ }\href {\doibase
  10.1103/PhysRevC.57.R1060} {\bibfield  {journal} {\bibinfo  {journal} {Phys.
  Rev. C}\ }\textbf {\bibinfo {volume} {57}},\ \bibinfo {pages} {R1060}
  (\bibinfo {year} {1998})}\BibitemShut {NoStop}%
\bibitem [{\citenamefont {Zhou}\ \emph {et~al.}(2008)\citenamefont {Zhou},
  \citenamefont {Polls}, \citenamefont {Schulze},\ and\ \citenamefont
  {Vidana}}]{Zhou2008_PRC78-054306}%
  \BibitemOpen
  \bibfield  {author} {\bibinfo {author} {\bibfnamefont {X.-R.}\ \bibnamefont
  {Zhou}}, \bibinfo {author} {\bibfnamefont {A.}~\bibnamefont {Polls}},
  \bibinfo {author} {\bibfnamefont {H.-J.}\ \bibnamefont {Schulze}}, \ and\
  \bibinfo {author} {\bibfnamefont {I.}~\bibnamefont {Vidana}},\ }\href
  {\doibase 10.1103/PhysRevC.78.054306} {\bibfield  {journal} {\bibinfo
  {journal} {Phys. Rev. C}\ }\textbf {\bibinfo {volume} {78}},\ \bibinfo
  {pages} {054306} (\bibinfo {year} {2008})}\BibitemShut {NoStop}%
\bibitem [{\citenamefont {Bohr}\ and\ \citenamefont
  {Mottelson}(1969)}]{Bohr1969_Nucl_Structure}%
  \BibitemOpen
  \bibfield  {author} {\bibinfo {author} {\bibfnamefont {A.}~\bibnamefont
  {Bohr}}\ and\ \bibinfo {author} {\bibfnamefont {B.~R.}\ \bibnamefont
  {Mottelson}},\ }\href@noop {} {\emph {\bibinfo {title} {Nuclear
  Structure}}},\ \bibinfo {edition} {1st}\ ed.,\ Vol.~\bibinfo {volume} {I}\
  (\bibinfo  {publisher} {Benjamin, New York},\ \bibinfo {year}
  {1969})\BibitemShut {NoStop}%
\bibitem [{\citenamefont {Paul}\ \emph {et~al.}(1988)\citenamefont {Paul},
  \citenamefont {Ma}, \citenamefont {Beausang}, \citenamefont {Fossan},
  \citenamefont {Piel}, \citenamefont {Shi}, \citenamefont {Xu},\ and\
  \citenamefont {Zhang}}]{Paul1988_PRL61-42}%
  \BibitemOpen
  \bibfield  {author} {\bibinfo {author} {\bibfnamefont {E.~S.}\ \bibnamefont
  {Paul}}, \bibinfo {author} {\bibfnamefont {R.}~\bibnamefont {Ma}}, \bibinfo
  {author} {\bibfnamefont {C.~W.}\ \bibnamefont {Beausang}}, \bibinfo {author}
  {\bibfnamefont {D.~B.}\ \bibnamefont {Fossan}}, \bibinfo {author}
  {\bibfnamefont {W.~F.}\ \bibnamefont {Piel}}, \bibinfo {author}
  {\bibfnamefont {S.}~\bibnamefont {Shi}}, \bibinfo {author} {\bibfnamefont
  {N.}~\bibnamefont {Xu}}, \ and\ \bibinfo {author} {\bibfnamefont {J.~y.}\
  \bibnamefont {Zhang}},\ }\href {\doibase 10.1103/PhysRevLett.61.42}
  {\bibfield  {journal} {\bibinfo  {journal} {Phys. Rev. Lett.}\ }\textbf
  {\bibinfo {volume} {61}},\ \bibinfo {pages} {42} (\bibinfo {year}
  {1988})}\BibitemShut {NoStop}%
\bibitem [{\citenamefont {Nazarewicz}\ \emph {et~al.}(1990)\citenamefont
  {Nazarewicz}, \citenamefont {Riley},\ and\ \citenamefont
  {Garrett}}]{Nazarewicz1990_NPA512-61}%
  \BibitemOpen
  \bibfield  {author} {\bibinfo {author} {\bibfnamefont {W.}~\bibnamefont
  {Nazarewicz}}, \bibinfo {author} {\bibfnamefont {M.~A.}\ \bibnamefont
  {Riley}}, \ and\ \bibinfo {author} {\bibfnamefont {J.~D.}\ \bibnamefont
  {Garrett}},\ }\href {\doibase 10.1016/0375-9474(90)90004-6} {\bibfield
  {journal} {\bibinfo  {journal} {Nucl. Phys. A}\ }\textbf {\bibinfo {volume}
  {512}},\ \bibinfo {pages} {61} (\bibinfo {year} {1990})}\BibitemShut
  {NoStop}%
\bibitem [{\citenamefont {Ma}\ \emph {et~al.}(1996{\natexlab{b}})\citenamefont
  {Ma}, \citenamefont {Zheng}, \citenamefont {Sun}, \citenamefont {Yang},
  \citenamefont {Zhou}, \citenamefont {Ding}, \citenamefont {Huo},
  \citenamefont {Liu}, \citenamefont {Liu}, \citenamefont {Wen},\ and\
  \citenamefont {Yang}}]{Ma1996_HEPNP20-865}%
  \BibitemOpen
  \bibfield  {author} {\bibinfo {author} {\bibfnamefont {Y.}~\bibnamefont
  {Ma}}, \bibinfo {author} {\bibfnamefont {H.}~\bibnamefont {Zheng}}, \bibinfo
  {author} {\bibfnamefont {H.}~\bibnamefont {Sun}}, \bibinfo {author}
  {\bibfnamefont {H.}~\bibnamefont {Yang}}, \bibinfo {author} {\bibfnamefont
  {S.}~\bibnamefont {Zhou}}, \bibinfo {author} {\bibfnamefont {Y.}~\bibnamefont
  {Ding}}, \bibinfo {author} {\bibfnamefont {J.}~\bibnamefont {Huo}}, \bibinfo
  {author} {\bibfnamefont {Y.}~\bibnamefont {Liu}}, \bibinfo {author}
  {\bibfnamefont {X.}~\bibnamefont {Liu}}, \bibinfo {author} {\bibfnamefont
  {S.}~\bibnamefont {Wen}}, \ and\ \bibinfo {author} {\bibfnamefont
  {C.}~\bibnamefont {Yang}},\ }\href
  {http://epub.cnki.net/grid2008/detail.aspx?filename=KNWL610.000&dbname=CJFD1996}
  {\bibfield  {journal} {\bibinfo  {journal} {High Ener. Phys. Nucl. Phys.}\
  }\textbf {\bibinfo {volume} {20}},\ \bibinfo {pages} {865} (\bibinfo {year}
  {1996}{\natexlab{b}})},\ \bibinfo {note} {in Chinese}\BibitemShut {NoStop}%
\bibitem [{\citenamefont {Zhou}\ \emph
  {et~al.}(2007{\natexlab{a}})\citenamefont {Zhou}, \citenamefont {Xing},
  \citenamefont {Liu}, \citenamefont {Zhang}, \citenamefont {Guo},
  \citenamefont {Ma}, \citenamefont {Lei}, \citenamefont {Guo}, \citenamefont
  {Oshima}, \citenamefont {Toh}, \citenamefont {Koizumi}, \citenamefont {Osa},
  \citenamefont {Hatsukawa}, \citenamefont {Xu},\ and\ \citenamefont
  {Sugawara}}]{Zhou2007_PRC75-034314}%
  \BibitemOpen
  \bibfield  {author} {\bibinfo {author} {\bibfnamefont {X.~H.}\ \bibnamefont
  {Zhou}}, \bibinfo {author} {\bibfnamefont {Y.~B.}\ \bibnamefont {Xing}},
  \bibinfo {author} {\bibfnamefont {M.~L.}\ \bibnamefont {Liu}}, \bibinfo
  {author} {\bibfnamefont {Y.~H.}\ \bibnamefont {Zhang}}, \bibinfo {author}
  {\bibfnamefont {Y.~X.}\ \bibnamefont {Guo}}, \bibinfo {author} {\bibfnamefont
  {L.}~\bibnamefont {Ma}}, \bibinfo {author} {\bibfnamefont {X.~G.}\
  \bibnamefont {Lei}}, \bibinfo {author} {\bibfnamefont {W.~T.}\ \bibnamefont
  {Guo}}, \bibinfo {author} {\bibfnamefont {M.}~\bibnamefont {Oshima}},
  \bibinfo {author} {\bibfnamefont {Y.}~\bibnamefont {Toh}}, \bibinfo {author}
  {\bibfnamefont {M.}~\bibnamefont {Koizumi}}, \bibinfo {author} {\bibfnamefont
  {A.}~\bibnamefont {Osa}}, \bibinfo {author} {\bibfnamefont {Y.}~\bibnamefont
  {Hatsukawa}}, \bibinfo {author} {\bibfnamefont {F.~R.}\ \bibnamefont {Xu}}, \
  and\ \bibinfo {author} {\bibfnamefont {M.}~\bibnamefont {Sugawara}},\ }\href
  {\doibase 10.1103/PhysRevC.75.034314} {\bibfield  {journal} {\bibinfo
  {journal} {Phys. Rev. C}\ }\textbf {\bibinfo {volume} {75}},\ \bibinfo
  {pages} {034314} (\bibinfo {year} {2007}{\natexlab{a}})}\BibitemShut
  {NoStop}%
\bibitem [{\citenamefont {Isaka}\ \emph {et~al.}(2011)\citenamefont {Isaka},
  \citenamefont {Kimura}, \citenamefont {Dote},\ and\ \citenamefont
  {Ohnishi}}]{Isaka2011_1104.3940}%
  \BibitemOpen
  \bibfield  {author} {\bibinfo {author} {\bibfnamefont {M.}~\bibnamefont
  {Isaka}}, \bibinfo {author} {\bibfnamefont {M.}~\bibnamefont {Kimura}},
  \bibinfo {author} {\bibfnamefont {A.}~\bibnamefont {Dote}}, \ and\ \bibinfo
  {author} {\bibfnamefont {A.}~\bibnamefont {Ohnishi}},\ }\href
  {http://arxiv.org/abs/1104.3940v1} {\enquote {\bibinfo {title} {Deformation
  of hypernuclei studied with antisymmetirzed molecular dynamics},}\ }\bibinfo
  {howpublished} {arXiv:1104.3940v1 [nucl-th]} (\bibinfo {year}
  {2011})\BibitemShut {NoStop}%
\bibitem [{\citenamefont {Rayet}(1976)}]{Rayet1976_AoP102-226}%
  \BibitemOpen
  \bibfield  {author} {\bibinfo {author} {\bibfnamefont {M.}~\bibnamefont
  {Rayet}},\ }\href {\doibase 10.1016/0003-4916(76)90262-1} {\bibfield
  {journal} {\bibinfo  {journal} {Ann. Phys.}\ }\textbf {\bibinfo {volume}
  {102}},\ \bibinfo {pages} {226} (\bibinfo {year} {1976})}\BibitemShut
  {NoStop}%
\bibitem [{\citenamefont {Rayet}(1981)}]{Rayet1981_NPA367-381}%
  \BibitemOpen
  \bibfield  {author} {\bibinfo {author} {\bibfnamefont {M.}~\bibnamefont
  {Rayet}},\ }\href {\doibase 10.1016/0375-9474(81)90655-2} {\bibfield
  {journal} {\bibinfo  {journal} {Nucl. Phys. A}\ }\textbf {\bibinfo {volume}
  {367}},\ \bibinfo {pages} {381} (\bibinfo {year} {1981})}\BibitemShut
  {NoStop}%
\bibitem [{\citenamefont {Lanskoy}\ and\ \citenamefont
  {Yamamoto}(1997)}]{Lanskoy1997_PRC55-2330}%
  \BibitemOpen
  \bibfield  {author} {\bibinfo {author} {\bibfnamefont {D.~E.}\ \bibnamefont
  {Lanskoy}}\ and\ \bibinfo {author} {\bibfnamefont {Y.}~\bibnamefont
  {Yamamoto}},\ }\href {\doibase 10.1103/PhysRevC.55.2330} {\bibfield
  {journal} {\bibinfo  {journal} {Phys. Rev. C}\ }\textbf {\bibinfo {volume}
  {55}},\ \bibinfo {pages} {2330} (\bibinfo {year} {1997})}\BibitemShut
  {NoStop}%
\bibitem [{\citenamefont {Lanskoy}(1998)}]{Lanskoy1998_PRC58-3351}%
  \BibitemOpen
  \bibfield  {author} {\bibinfo {author} {\bibfnamefont {D.~E.}\ \bibnamefont
  {Lanskoy}},\ }\href {\doibase 10.1103/PhysRevC.58.3351} {\bibfield  {journal}
  {\bibinfo  {journal} {Phys. Rev. C}\ }\textbf {\bibinfo {volume} {58}},\
  \bibinfo {pages} {3351} (\bibinfo {year} {1998})}\BibitemShut {NoStop}%
\bibitem [{\citenamefont {Cugnon}\ \emph {et~al.}(2000)\citenamefont {Cugnon},
  \citenamefont {Lejeune},\ and\ \citenamefont
  {Schulze}}]{Cugnon2000_PRC62-064308}%
  \BibitemOpen
  \bibfield  {author} {\bibinfo {author} {\bibfnamefont {J.}~\bibnamefont
  {Cugnon}}, \bibinfo {author} {\bibfnamefont {A.}~\bibnamefont {Lejeune}}, \
  and\ \bibinfo {author} {\bibfnamefont {H.-J.}\ \bibnamefont {Schulze}},\
  }\href {\doibase 10.1103/PhysRevC.62.064308} {\bibfield  {journal} {\bibinfo
  {journal} {Phys. Rev. C}\ }\textbf {\bibinfo {volume} {62}},\ \bibinfo
  {pages} {064308} (\bibinfo {year} {2000})}\BibitemShut {NoStop}%
\bibitem [{\citenamefont {Vidana}\ \emph {et~al.}(2001)\citenamefont {Vidana},
  \citenamefont {Polls}, \citenamefont {Ramos},\ and\ \citenamefont
  {Schulze}}]{Vidana2001_PRC64-044301}%
  \BibitemOpen
  \bibfield  {author} {\bibinfo {author} {\bibfnamefont {I.}~\bibnamefont
  {Vidana}}, \bibinfo {author} {\bibfnamefont {A.}~\bibnamefont {Polls}},
  \bibinfo {author} {\bibfnamefont {A.}~\bibnamefont {Ramos}}, \ and\ \bibinfo
  {author} {\bibfnamefont {H.-J.}\ \bibnamefont {Schulze}},\ }\href {\doibase
  10.1103/PhysRevC.64.044301} {\bibfield  {journal} {\bibinfo  {journal} {Phys.
  Rev. C}\ }\textbf {\bibinfo {volume} {64}},\ \bibinfo {pages} {044301}
  (\bibinfo {year} {2001})}\BibitemShut {NoStop}%
\bibitem [{\citenamefont {Brockmann}\ and\ \citenamefont
  {Weise}(1977)}]{Brockmann1977_PLB69-167}%
  \BibitemOpen
  \bibfield  {author} {\bibinfo {author} {\bibfnamefont {R.}~\bibnamefont
  {Brockmann}}\ and\ \bibinfo {author} {\bibfnamefont {W.}~\bibnamefont
  {Weise}},\ }\href {\doibase 10.1016/0370-2693(77)90635-9} {\bibfield
  {journal} {\bibinfo  {journal} {Phys. Lett. B}\ }\textbf {\bibinfo {volume}
  {69}},\ \bibinfo {pages} {167} (\bibinfo {year} {1977})}\BibitemShut
  {NoStop}%
\bibitem [{\citenamefont {Bouyssy}(1982)}]{Bouyssy1982_NPA381-445}%
  \BibitemOpen
  \bibfield  {author} {\bibinfo {author} {\bibfnamefont {A.}~\bibnamefont
  {Bouyssy}},\ }\href {\doibase 10.1016/0375-9474(82)90370-0} {\bibfield
  {journal} {\bibinfo  {journal} {Nucl. Phys. A}\ }\textbf {\bibinfo {volume}
  {381}},\ \bibinfo {pages} {445} (\bibinfo {year} {1982})}\BibitemShut
  {NoStop}%
\bibitem [{\citenamefont {Mares}\ and\ \citenamefont
  {Zofka}(1989)}]{Mares1989_ZPA333-209}%
  \BibitemOpen
  \bibfield  {author} {\bibinfo {author} {\bibfnamefont {J.}~\bibnamefont
  {Mares}}\ and\ \bibinfo {author} {\bibfnamefont {J.}~\bibnamefont {Zofka}},\
  }\href {\doibase 10.1007/BF01565152} {\bibfield  {journal} {\bibinfo
  {journal} {Z. Phys. A}\ }\textbf {\bibinfo {volume} {333}},\ \bibinfo {pages}
  {209} (\bibinfo {year} {1989})}\BibitemShut {NoStop}%
\bibitem [{\citenamefont {Mares}\ and\ \citenamefont
  {Zofka}(1990)}]{Mares1990_PLB249-181}%
  \BibitemOpen
  \bibfield  {author} {\bibinfo {author} {\bibfnamefont {J.}~\bibnamefont
  {Mares}}\ and\ \bibinfo {author} {\bibfnamefont {J.}~\bibnamefont {Zofka}},\
  }\href {\doibase 10.1016/0370-2693(90)91239-8} {\bibfield  {journal}
  {\bibinfo  {journal} {Phys. Lett. B}\ }\textbf {\bibinfo {volume} {249}},\
  \bibinfo {pages} {181} (\bibinfo {year} {1990})}\BibitemShut {NoStop}%
\bibitem [{\citenamefont {Rufa}\ \emph {et~al.}(1990)\citenamefont {Rufa},
  \citenamefont {Schaffner}, \citenamefont {Maruhn}, \citenamefont {Stoecker},
  \citenamefont {Greiner},\ and\ \citenamefont
  {Reinhard}}]{Rufa1990_PRC42-2469}%
  \BibitemOpen
  \bibfield  {author} {\bibinfo {author} {\bibfnamefont {M.}~\bibnamefont
  {Rufa}}, \bibinfo {author} {\bibfnamefont {J.}~\bibnamefont {Schaffner}},
  \bibinfo {author} {\bibfnamefont {J.}~\bibnamefont {Maruhn}}, \bibinfo
  {author} {\bibfnamefont {H.}~\bibnamefont {Stoecker}}, \bibinfo {author}
  {\bibfnamefont {W.}~\bibnamefont {Greiner}}, \ and\ \bibinfo {author}
  {\bibfnamefont {P.-G.}\ \bibnamefont {Reinhard}},\ }\href {\doibase
  10.1103/PhysRevC.42.2469} {\bibfield  {journal} {\bibinfo  {journal} {Phys.
  Rev. C}\ }\textbf {\bibinfo {volume} {42}},\ \bibinfo {pages} {2469}
  (\bibinfo {year} {1990})}\BibitemShut {NoStop}%
\bibitem [{\citenamefont {Cohen}\ and\ \citenamefont
  {Weber}(1991)}]{Cohen1991_PRC44-1181}%
  \BibitemOpen
  \bibfield  {author} {\bibinfo {author} {\bibfnamefont {J.}~\bibnamefont
  {Cohen}}\ and\ \bibinfo {author} {\bibfnamefont {H.~J.}\ \bibnamefont
  {Weber}},\ }\href {\doibase 10.1103/PhysRevC.44.1181} {\bibfield  {journal}
  {\bibinfo  {journal} {Phys. Rev. C}\ }\textbf {\bibinfo {volume} {44}},\
  \bibinfo {pages} {1181} (\bibinfo {year} {1991})}\BibitemShut {NoStop}%
\bibitem [{\citenamefont {Glendenning}\ and\ \citenamefont
  {Moszkowski}(1991)}]{Glendenning1991_PRL67-2414}%
  \BibitemOpen
  \bibfield  {author} {\bibinfo {author} {\bibfnamefont {N.~K.}\ \bibnamefont
  {Glendenning}}\ and\ \bibinfo {author} {\bibfnamefont {S.~A.}\ \bibnamefont
  {Moszkowski}},\ }\href {\doibase 10.1103/PhysRevLett.67.2414} {\bibfield
  {journal} {\bibinfo  {journal} {Phys. Rev. Lett.}\ }\textbf {\bibinfo
  {volume} {67}},\ \bibinfo {pages} {2414} (\bibinfo {year}
  {1991})}\BibitemShut {NoStop}%
\bibitem [{\citenamefont {Glendenning}\ \emph {et~al.}(1993)\citenamefont
  {Glendenning}, \citenamefont {Von-Eiff}, \citenamefont {Haft}, \citenamefont
  {Lenske},\ and\ \citenamefont {Weigel}}]{Glendenning1993_PRC48-889}%
  \BibitemOpen
  \bibfield  {author} {\bibinfo {author} {\bibfnamefont {N.~K.}\ \bibnamefont
  {Glendenning}}, \bibinfo {author} {\bibfnamefont {D.}~\bibnamefont
  {Von-Eiff}}, \bibinfo {author} {\bibfnamefont {M.}~\bibnamefont {Haft}},
  \bibinfo {author} {\bibfnamefont {H.}~\bibnamefont {Lenske}}, \ and\ \bibinfo
  {author} {\bibfnamefont {M.~K.}\ \bibnamefont {Weigel}},\ }\href {\doibase
  10.1103/PhysRevC.48.889} {\bibfield  {journal} {\bibinfo  {journal} {Phys.
  Rev. C}\ }\textbf {\bibinfo {volume} {48}},\ \bibinfo {pages} {889} (\bibinfo
  {year} {1993})}\BibitemShut {NoStop}%
\bibitem [{\citenamefont {Sugahara}\ and\ \citenamefont
  {Toki}(1994)}]{Sugahara1994_PTP92-803}%
  \BibitemOpen
  \bibfield  {author} {\bibinfo {author} {\bibfnamefont {Y.}~\bibnamefont
  {Sugahara}}\ and\ \bibinfo {author} {\bibfnamefont {H.}~\bibnamefont
  {Toki}},\ }\href {\doibase 10.1143/PTP.92.803} {\bibfield  {journal}
  {\bibinfo  {journal} {Prog. Theo. Phys.}\ }\textbf {\bibinfo {volume} {92}},\
  \bibinfo {pages} {803} (\bibinfo {year} {1994})}\BibitemShut {NoStop}%
\bibitem [{\citenamefont {Mares}\ and\ \citenamefont
  {Jennings}(1994)}]{Mares1994_PRC49-2472}%
  \BibitemOpen
  \bibfield  {author} {\bibinfo {author} {\bibfnamefont {J.}~\bibnamefont
  {Mares}}\ and\ \bibinfo {author} {\bibfnamefont {B.~K.}\ \bibnamefont
  {Jennings}},\ }\href {\doibase 10.1103/PhysRevC.49.2472} {\bibfield
  {journal} {\bibinfo  {journal} {Phys. Rev. C}\ }\textbf {\bibinfo {volume}
  {49}},\ \bibinfo {pages} {2472} (\bibinfo {year} {1994})}\BibitemShut
  {NoStop}%
\bibitem [{\citenamefont {Marcos}\ \emph {et~al.}(1998)\citenamefont {Marcos},
  \citenamefont {Lombard},\ and\ \citenamefont
  {Marescaron}}]{Marcos1998_PRC57-1178}%
  \BibitemOpen
  \bibfield  {author} {\bibinfo {author} {\bibfnamefont {S.}~\bibnamefont
  {Marcos}}, \bibinfo {author} {\bibfnamefont {R.~J.}\ \bibnamefont {Lombard}},
  \ and\ \bibinfo {author} {\bibfnamefont {J.}~\bibnamefont {Marescaron}},\
  }\href {\doibase 10.1103/PhysRevC.57.1178} {\bibfield  {journal} {\bibinfo
  {journal} {Phys. Rev. C}\ }\textbf {\bibinfo {volume} {57}},\ \bibinfo
  {pages} {1178} (\bibinfo {year} {1998})}\BibitemShut {NoStop}%
\bibitem [{\citenamefont {Keil}\ and\ \citenamefont
  {Lenske}(2002)}]{Keil2002_PRC66-054307}%
  \BibitemOpen
  \bibfield  {author} {\bibinfo {author} {\bibfnamefont {C.}~\bibnamefont
  {Keil}}\ and\ \bibinfo {author} {\bibfnamefont {H.}~\bibnamefont {Lenske}},\
  }\href {\doibase 10.1103/PhysRevC.66.054307} {\bibfield  {journal} {\bibinfo
  {journal} {Phys. Rev. C}\ }\textbf {\bibinfo {volume} {66}},\ \bibinfo
  {pages} {054307} (\bibinfo {year} {2002})}\BibitemShut {NoStop}%
\bibitem [{\citenamefont {Shen}\ \emph {et~al.}(2006)\citenamefont {Shen},
  \citenamefont {Yang},\ and\ \citenamefont {Toki}}]{Shen2006_PTP115-325}%
  \BibitemOpen
  \bibfield  {author} {\bibinfo {author} {\bibfnamefont {H.}~\bibnamefont
  {Shen}}, \bibinfo {author} {\bibfnamefont {F.}~\bibnamefont {Yang}}, \ and\
  \bibinfo {author} {\bibfnamefont {H.}~\bibnamefont {Toki}},\ }\href {\doibase
  10.1143/PTP.115.325} {\bibfield  {journal} {\bibinfo  {journal} {Prog. Theo.
  Phys.}\ }\textbf {\bibinfo {volume} {115}},\ \bibinfo {pages} {325} (\bibinfo
  {year} {2006})}\BibitemShut {NoStop}%
\bibitem [{\citenamefont {Zhong}\ \emph {et~al.}(2006)\citenamefont {Zhong},
  \citenamefont {Peng}, \citenamefont {Li},\ and\ \citenamefont
  {Ning}}]{Zhong2006_PRC74-034321}%
  \BibitemOpen
  \bibfield  {author} {\bibinfo {author} {\bibfnamefont {X.~H.}\ \bibnamefont
  {Zhong}}, \bibinfo {author} {\bibfnamefont {G.~X.}\ \bibnamefont {Peng}},
  \bibinfo {author} {\bibfnamefont {L.}~\bibnamefont {Li}}, \ and\ \bibinfo
  {author} {\bibfnamefont {P.~Z.}\ \bibnamefont {Ning}},\ }\href {\doibase
  10.1103/PhysRevC.74.034321} {\bibfield  {journal} {\bibinfo  {journal} {Phys.
  Rev. C}\ }\textbf {\bibinfo {volume} {74}},\ \bibinfo {pages} {034321}
  (\bibinfo {year} {2006})}\BibitemShut {NoStop}%
\bibitem [{\citenamefont {Song}\ \emph {et~al.}(2009)\citenamefont {Song},
  \citenamefont {Yao},\ and\ \citenamefont {Meng}}]{Song2009_CPL26-122102}%
  \BibitemOpen
  \bibfield  {author} {\bibinfo {author} {\bibfnamefont {C.-Y.}\ \bibnamefont
  {Song}}, \bibinfo {author} {\bibfnamefont {J.-M.}\ \bibnamefont {Yao}}, \
  and\ \bibinfo {author} {\bibfnamefont {J.}~\bibnamefont {Meng}},\ }\href
  {\doibase 10.1088/0256-307X/26/12/122102} {\bibfield  {journal} {\bibinfo
  {journal} {Chin. Phys. Lett.}\ }\textbf {\bibinfo {volume} {26}},\ \bibinfo
  {pages} {122102} (\bibinfo {year} {2009})}\BibitemShut {NoStop}%
\bibitem [{\citenamefont {Zhou}\ \emph
  {et~al.}(2007{\natexlab{b}})\citenamefont {Zhou}, \citenamefont {Schulze},
  \citenamefont {Sagawa}, \citenamefont {Wu},\ and\ \citenamefont
  {Zhao}}]{Zhou2007_PRC76-034312}%
  \BibitemOpen
  \bibfield  {author} {\bibinfo {author} {\bibfnamefont {X.-R.}\ \bibnamefont
  {Zhou}}, \bibinfo {author} {\bibfnamefont {H.-J.}\ \bibnamefont {Schulze}},
  \bibinfo {author} {\bibfnamefont {H.}~\bibnamefont {Sagawa}}, \bibinfo
  {author} {\bibfnamefont {C.-X.}\ \bibnamefont {Wu}}, \ and\ \bibinfo {author}
  {\bibfnamefont {E.-G.}\ \bibnamefont {Zhao}},\ }\href {\doibase
  10.1103/PhysRevC.76.034312} {\bibfield  {journal} {\bibinfo  {journal} {Phys.
  Rev. C}\ }\textbf {\bibinfo {volume} {76}},\ \bibinfo {pages} {034312}
  (\bibinfo {year} {2007}{\natexlab{b}})}\BibitemShut {NoStop}%
\bibitem [{\citenamefont {Zhou}\ \emph {et~al.}(2009)\citenamefont {Zhou},
  \citenamefont {Cui},\ and\ \citenamefont {Wei}}]{Zhou2009_SCG52-1548}%
  \BibitemOpen
  \bibfield  {author} {\bibinfo {author} {\bibfnamefont {X.-R.}\ \bibnamefont
  {Zhou}}, \bibinfo {author} {\bibfnamefont {J.-W.}\ \bibnamefont {Cui}}, \
  and\ \bibinfo {author} {\bibfnamefont {N.}~\bibnamefont {Wei}},\ }\href
  {\doibase 10.1007/s11433-009-0212-0} {\bibfield  {journal} {\bibinfo
  {journal} {Sci. China G}\ }\textbf {\bibinfo {volume} {52}},\ \bibinfo
  {pages} {1548} (\bibinfo {year} {2009})}\BibitemShut {NoStop}%
\bibitem [{\citenamefont {Win}\ and\ \citenamefont
  {Hagino}(2008)}]{Win2008_PRC78-054311}%
  \BibitemOpen
  \bibfield  {author} {\bibinfo {author} {\bibfnamefont {M.~T.}\ \bibnamefont
  {Win}}\ and\ \bibinfo {author} {\bibfnamefont {K.}~\bibnamefont {Hagino}},\
  }\href {\doibase 10.1103/PhysRevC.78.054311} {\bibfield  {journal} {\bibinfo
  {journal} {Phys. Rev. C}\ }\textbf {\bibinfo {volume} {78}},\ \bibinfo
  {pages} {054311} (\bibinfo {year} {2008})}\BibitemShut {NoStop}%
\bibitem [{\citenamefont {Schulze}\ \emph {et~al.}(2010)\citenamefont
  {Schulze}, \citenamefont {Win}, \citenamefont {Hagino},\ and\ \citenamefont
  {Sagawa}}]{Schulze2010_PTP123-569}%
  \BibitemOpen
  \bibfield  {author} {\bibinfo {author} {\bibfnamefont {H.-J.}\ \bibnamefont
  {Schulze}}, \bibinfo {author} {\bibfnamefont {M.~T.}\ \bibnamefont {Win}},
  \bibinfo {author} {\bibfnamefont {K.}~\bibnamefont {Hagino}}, \ and\ \bibinfo
  {author} {\bibfnamefont {H.~S.}\ \bibnamefont {Sagawa}},\ }\href {\doibase
  10.1143/PTP.123.569} {\bibfield  {journal} {\bibinfo  {journal} {Prog. Theo.
  Phys.}\ }\textbf {\bibinfo {volume} {123}},\ \bibinfo {pages} {569} (\bibinfo
  {year} {2010})}\BibitemShut {NoStop}%
\bibitem [{\citenamefont {Frauendorf}\ and\ \citenamefont
  {Meng}(1997)}]{Frauendorf1997_NPA617-131}%
  \BibitemOpen
  \bibfield  {author} {\bibinfo {author} {\bibfnamefont {S.}~\bibnamefont
  {Frauendorf}}\ and\ \bibinfo {author} {\bibfnamefont {J.}~\bibnamefont
  {Meng}},\ }\href {\doibase 10.1016/S0375-9474(97)00004-3} {\bibfield
  {journal} {\bibinfo  {journal} {Nucl. Phys. A}\ }\textbf {\bibinfo {volume}
  {617}},\ \bibinfo {pages} {131} (\bibinfo {year} {1997})}\BibitemShut
  {NoStop}%
\bibitem [{\citenamefont {Frauendorf}(2001)}]{Frauendorf2001_RMP73-463}%
  \BibitemOpen
  \bibfield  {author} {\bibinfo {author} {\bibfnamefont {S.}~\bibnamefont
  {Frauendorf}},\ }\href {\doibase 10.1103/RevModPhys.73.463} {\bibfield
  {journal} {\bibinfo  {journal} {Rev. Mod. Phys.}\ }\textbf {\bibinfo {volume}
  {73}},\ \bibinfo {pages} {463} (\bibinfo {year} {2001})}\BibitemShut
  {NoStop}%
\bibitem [{\citenamefont {Meng}\ and\ \citenamefont
  {Zhang}(2010)}]{Meng2010_JPG37-064025}%
  \BibitemOpen
  \bibfield  {author} {\bibinfo {author} {\bibfnamefont {J.}~\bibnamefont
  {Meng}}\ and\ \bibinfo {author} {\bibfnamefont {S.~Q.}\ \bibnamefont
  {Zhang}},\ }\href {\doibase 10.1088/0954-3899/37/6/064025} {\bibfield
  {journal} {\bibinfo  {journal} {J. Phys. G: Nucl. Phys.}\ }\textbf {\bibinfo
  {volume} {37}},\ \bibinfo {pages} {064025} (\bibinfo {year}
  {2010})}\BibitemShut {NoStop}%
\bibitem [{\citenamefont {Odegard}\ \emph {et~al.}(2001)\citenamefont
  {Odegard}, \citenamefont {Hagemann}, \citenamefont {Jensen}, \citenamefont
  {Bergstroem}, \citenamefont {Herskind}, \citenamefont {Sletten},
  \citenamefont {Toermaenen}, \citenamefont {Wilson}, \citenamefont {Tjom},
  \citenamefont {Hamamoto}, \citenamefont {Spohr}, \citenamefont {Huebel},
  \citenamefont {Goergen}, \citenamefont {Schoenwasser}, \citenamefont
  {Bracco}, \citenamefont {Leoni}, \citenamefont {Maj}, \citenamefont
  {Petrache}, \citenamefont {Bednarczyk},\ and\ \citenamefont
  {Curien}}]{Odegard2001_PRL86-5866}%
  \BibitemOpen
  \bibfield  {author} {\bibinfo {author} {\bibfnamefont {S.~W.}\ \bibnamefont
  {Odegard}}, \bibinfo {author} {\bibfnamefont {G.~B.}\ \bibnamefont
  {Hagemann}}, \bibinfo {author} {\bibfnamefont {D.~R.}\ \bibnamefont
  {Jensen}}, \bibinfo {author} {\bibfnamefont {M.}~\bibnamefont {Bergstroem}},
  \bibinfo {author} {\bibfnamefont {B.}~\bibnamefont {Herskind}}, \bibinfo
  {author} {\bibfnamefont {G.}~\bibnamefont {Sletten}}, \bibinfo {author}
  {\bibfnamefont {S.}~\bibnamefont {Toermaenen}}, \bibinfo {author}
  {\bibfnamefont {J.~N.}\ \bibnamefont {Wilson}}, \bibinfo {author}
  {\bibfnamefont {P.~O.}\ \bibnamefont {Tjom}}, \bibinfo {author}
  {\bibfnamefont {I.}~\bibnamefont {Hamamoto}}, \bibinfo {author}
  {\bibfnamefont {K.}~\bibnamefont {Spohr}}, \bibinfo {author} {\bibfnamefont
  {H.}~\bibnamefont {Huebel}}, \bibinfo {author} {\bibfnamefont
  {A.}~\bibnamefont {Goergen}}, \bibinfo {author} {\bibfnamefont
  {G.}~\bibnamefont {Schoenwasser}}, \bibinfo {author} {\bibfnamefont
  {A.}~\bibnamefont {Bracco}}, \bibinfo {author} {\bibfnamefont
  {S.}~\bibnamefont {Leoni}}, \bibinfo {author} {\bibfnamefont
  {A.}~\bibnamefont {Maj}}, \bibinfo {author} {\bibfnamefont {C.~M.}\
  \bibnamefont {Petrache}}, \bibinfo {author} {\bibfnamefont {P.}~\bibnamefont
  {Bednarczyk}}, \ and\ \bibinfo {author} {\bibfnamefont {D.}~\bibnamefont
  {Curien}},\ }\href {\doibase 10.1103/PhysRevLett.86.5866} {\bibfield
  {journal} {\bibinfo  {journal} {Phys. Rev. Lett.}\ }\textbf {\bibinfo
  {volume} {86}},\ \bibinfo {pages} {5866} (\bibinfo {year}
  {2001})}\BibitemShut {NoStop}%
\bibitem [{\citenamefont {Chen}\ and\ \citenamefont
  {Gao}(2011)}]{Chen2011_NSC2010}%
  \BibitemOpen
  \bibfield  {author} {\bibinfo {author} {\bibfnamefont {Y.~S.}\ \bibnamefont
  {Chen}}\ and\ \bibinfo {author} {\bibfnamefont {Z.-C.}\ \bibnamefont {Gao}},\
  }in\ \href@noop {} {\emph {\bibinfo {booktitle} {Nuclear Structure in China
  2010: Proceedings of the 13th National Conference on Nuclear Structure in
  China (NSC2010)}}}\ (\bibinfo  {publisher} {World Scientific},\ \bibinfo
  {year} {2011})\BibitemShut {NoStop}%
\bibitem [{\citenamefont {Guo}\ \emph {et~al.}(2007{\natexlab{a}})\citenamefont
  {Guo}, \citenamefont {Maruhn},\ and\ \citenamefont
  {Reinhard}}]{Guo2007_PRC76-034317}%
  \BibitemOpen
  \bibfield  {author} {\bibinfo {author} {\bibfnamefont {L.}~\bibnamefont
  {Guo}}, \bibinfo {author} {\bibfnamefont {J.~A.}\ \bibnamefont {Maruhn}}, \
  and\ \bibinfo {author} {\bibfnamefont {P.-G.}\ \bibnamefont {Reinhard}},\
  }\href {\doibase 10.1103/PhysRevC.76.034317} {\bibfield  {journal} {\bibinfo
  {journal} {Phys. Rev. C}\ }\textbf {\bibinfo {volume} {76}},\ \bibinfo
  {pages} {034317} (\bibinfo {year} {2007}{\natexlab{a}})}\BibitemShut
  {NoStop}%
\bibitem [{\citenamefont {Guo}\ \emph {et~al.}(2007{\natexlab{b}})\citenamefont
  {Guo}, \citenamefont {Hempel}, \citenamefont {Schaffner-Bielich},\ and\
  \citenamefont {Maruhn}}]{Guo2007_PRC76-065801}%
  \BibitemOpen
  \bibfield  {author} {\bibinfo {author} {\bibfnamefont {L.}~\bibnamefont
  {Guo}}, \bibinfo {author} {\bibfnamefont {M.}~\bibnamefont {Hempel}},
  \bibinfo {author} {\bibfnamefont {J.}~\bibnamefont {Schaffner-Bielich}}, \
  and\ \bibinfo {author} {\bibfnamefont {J.~A.}\ \bibnamefont {Maruhn}},\
  }\href {\doibase 10.1103/PhysRevC.76.065801} {\bibfield  {journal} {\bibinfo
  {journal} {Phys. Rev. C}\ }\textbf {\bibinfo {volume} {76}},\ \bibinfo
  {pages} {065801} (\bibinfo {year} {2007}{\natexlab{b}})}\BibitemShut
  {NoStop}%
\bibitem [{\citenamefont {Li}\ \emph {et~al.}(2009)\citenamefont {Li},
  \citenamefont {Niksic}, \citenamefont {Vretenar},\ and\ \citenamefont
  {Meng}}]{Li2009_PRC80-061301}%
  \BibitemOpen
  \bibfield  {author} {\bibinfo {author} {\bibfnamefont {Z.~P.}\ \bibnamefont
  {Li}}, \bibinfo {author} {\bibfnamefont {T.}~\bibnamefont {Niksic}}, \bibinfo
  {author} {\bibfnamefont {D.}~\bibnamefont {Vretenar}}, \ and\ \bibinfo
  {author} {\bibfnamefont {J.}~\bibnamefont {Meng}},\ }\href {\doibase
  10.1103/PhysRevC.80.061301} {\bibfield  {journal} {\bibinfo  {journal} {Phys.
  Rev. C}\ }\textbf {\bibinfo {volume} {80}},\ \bibinfo {pages} {061301}
  (\bibinfo {year} {2009})}\BibitemShut {NoStop}%
\bibitem [{\citenamefont {Pashkevich}(1969)}]{Pashkevich1969_NPA133-400}%
  \BibitemOpen
  \bibfield  {author} {\bibinfo {author} {\bibfnamefont {V.~V.}\ \bibnamefont
  {Pashkevich}},\ }\href {\doibase 10.1016/0375-9474(69)90641-1} {\bibfield
  {journal} {\bibinfo  {journal} {Nucl. Phys. A}\ }\textbf {\bibinfo {volume}
  {133}},\ \bibinfo {pages} {400} (\bibinfo {year} {1969})}\BibitemShut
  {NoStop}%
\bibitem [{\citenamefont {Abusara}\ \emph {et~al.}(2010)\citenamefont
  {Abusara}, \citenamefont {Afanasjev},\ and\ \citenamefont
  {Ring}}]{Abusara2010_PRC82-044303}%
  \BibitemOpen
  \bibfield  {author} {\bibinfo {author} {\bibfnamefont {H.}~\bibnamefont
  {Abusara}}, \bibinfo {author} {\bibfnamefont {A.~V.}\ \bibnamefont
  {Afanasjev}}, \ and\ \bibinfo {author} {\bibfnamefont {P.}~\bibnamefont
  {Ring}},\ }\href {\doibase 10.1103/PhysRevC.82.044303} {\bibfield  {journal}
  {\bibinfo  {journal} {Phys. Rev. C}\ }\textbf {\bibinfo {volume} {82}},\
  \bibinfo {pages} {044303} (\bibinfo {year} {2010})}\BibitemShut {NoStop}%
\bibitem [{\citenamefont {Win}\ \emph {et~al.}(2011)\citenamefont {Win},
  \citenamefont {Hagino},\ and\ \citenamefont {Koike}}]{Win2011_PRC83-014301}%
  \BibitemOpen
  \bibfield  {author} {\bibinfo {author} {\bibfnamefont {M.~T.}\ \bibnamefont
  {Win}}, \bibinfo {author} {\bibfnamefont {K.}~\bibnamefont {Hagino}}, \ and\
  \bibinfo {author} {\bibfnamefont {T.}~\bibnamefont {Koike}},\ }\href
  {\doibase 10.1103/PhysRevC.83.014301} {\bibfield  {journal} {\bibinfo
  {journal} {Phys. Rev. C}\ }\textbf {\bibinfo {volume} {83}},\ \bibinfo
  {pages} {014301} (\bibinfo {year} {2011})}\BibitemShut {NoStop}%
\bibitem [{\citenamefont {Serot}\ and\ \citenamefont
  {Walecka}(1986)}]{Serot1986_ANP16-1}%
  \BibitemOpen
  \bibfield  {author} {\bibinfo {author} {\bibfnamefont {B.~D.}\ \bibnamefont
  {Serot}}\ and\ \bibinfo {author} {\bibfnamefont {J.~D.}\ \bibnamefont
  {Walecka}},\ }\href@noop {} {\bibfield  {journal} {\bibinfo  {journal} {Adv.
  Nucl. Phys.}\ }\textbf {\bibinfo {volume} {16}},\ \bibinfo {pages} {1}
  (\bibinfo {year} {1986})}\BibitemShut {NoStop}%
\bibitem [{\citenamefont {Reinhard}(1989)}]{Reinhard1989_RPP52-439}%
  \BibitemOpen
  \bibfield  {author} {\bibinfo {author} {\bibfnamefont {P.~G.}\ \bibnamefont
  {Reinhard}},\ }\href {\doibase 10.1088/0034-4885/52/4/002} {\bibfield
  {journal} {\bibinfo  {journal} {Rep. Prog. Phys.}\ }\textbf {\bibinfo
  {volume} {52}},\ \bibinfo {pages} {439} (\bibinfo {year} {1989})}\BibitemShut
  {NoStop}%
\bibitem [{\citenamefont {Ring}(1996)}]{Ring1996_PPNP37-193}%
  \BibitemOpen
  \bibfield  {author} {\bibinfo {author} {\bibfnamefont {P.}~\bibnamefont
  {Ring}},\ }\href {\doibase 10.1016/0146-6410(96)00054-3} {\bibfield
  {journal} {\bibinfo  {journal} {Prog. Part. Nucl. Phys.}\ }\textbf {\bibinfo
  {volume} {37}},\ \bibinfo {pages} {193} (\bibinfo {year} {1996})}\BibitemShut
  {NoStop}%
\bibitem [{\citenamefont {Vretenar}\ \emph {et~al.}(2005)\citenamefont
  {Vretenar}, \citenamefont {Afanasjev}, \citenamefont {Lalazissis},\ and\
  \citenamefont {Ring}}]{Vretenar2005_PR409-101}%
  \BibitemOpen
  \bibfield  {author} {\bibinfo {author} {\bibfnamefont {D.}~\bibnamefont
  {Vretenar}}, \bibinfo {author} {\bibfnamefont {A.}~\bibnamefont {Afanasjev}},
  \bibinfo {author} {\bibfnamefont {G.}~\bibnamefont {Lalazissis}}, \ and\
  \bibinfo {author} {\bibfnamefont {P.}~\bibnamefont {Ring}},\ }\href {\doibase
  10.1016/j.physrep.2004.10.001} {\bibfield  {journal} {\bibinfo  {journal}
  {Phys. Rep.}\ }\textbf {\bibinfo {volume} {409}},\ \bibinfo {pages} {101}
  (\bibinfo {year} {2005})}\BibitemShut {NoStop}%
\bibitem [{\citenamefont {Meng}\ \emph
  {et~al.}(2006{\natexlab{a}})\citenamefont {Meng}, \citenamefont {Toki},
  \citenamefont {Zhou}, \citenamefont {Zhang}, \citenamefont {Long},\ and\
  \citenamefont {Geng}}]{Meng2006_PPNP57-470}%
  \BibitemOpen
  \bibfield  {author} {\bibinfo {author} {\bibfnamefont {J.}~\bibnamefont
  {Meng}}, \bibinfo {author} {\bibfnamefont {H.}~\bibnamefont {Toki}}, \bibinfo
  {author} {\bibfnamefont {S.~G.}\ \bibnamefont {Zhou}}, \bibinfo {author}
  {\bibfnamefont {S.~Q.}\ \bibnamefont {Zhang}}, \bibinfo {author}
  {\bibfnamefont {W.~H.}\ \bibnamefont {Long}}, \ and\ \bibinfo {author}
  {\bibfnamefont {L.~S.}\ \bibnamefont {Geng}},\ }\href {\doibase
  10.1016/j.ppnp.2005.06.001} {\bibfield  {journal} {\bibinfo  {journal} {Prog.
  Part. Nucl. Phys.}\ }\textbf {\bibinfo {volume} {57}},\ \bibinfo {pages}
  {470} (\bibinfo {year} {2006}{\natexlab{a}})}\BibitemShut {NoStop}%
\bibitem [{\citenamefont {Jennings}(1990)}]{Jennings1990_PLB246-325}%
  \BibitemOpen
  \bibfield  {author} {\bibinfo {author} {\bibfnamefont {B.~K.}\ \bibnamefont
  {Jennings}},\ }\href {\doibase 10.1016/0370-2693(90)90607-8} {\bibfield
  {journal} {\bibinfo  {journal} {Phys. Lett. B}\ }\textbf {\bibinfo {volume}
  {246}},\ \bibinfo {pages} {325} (\bibinfo {year} {1990})}\BibitemShut
  {NoStop}%
\bibitem [{\citenamefont {Hirata}\ \emph {et~al.}(1996)\citenamefont {Hirata},
  \citenamefont {Sumiyoshi}, \citenamefont {Carlson}, \citenamefont {Toki},\
  and\ \citenamefont {Tanihata}}]{Hirata1996_NPA609-131}%
  \BibitemOpen
  \bibfield  {author} {\bibinfo {author} {\bibfnamefont {D.}~\bibnamefont
  {Hirata}}, \bibinfo {author} {\bibfnamefont {K.}~\bibnamefont {Sumiyoshi}},
  \bibinfo {author} {\bibfnamefont {B.~V.}\ \bibnamefont {Carlson}}, \bibinfo
  {author} {\bibfnamefont {H.}~\bibnamefont {Toki}}, \ and\ \bibinfo {author}
  {\bibfnamefont {I.}~\bibnamefont {Tanihata}},\ }\href {\doibase
  10.1016/0375-9474(96)00298-9} {\bibfield  {journal} {\bibinfo  {journal}
  {Nucl. Phys. A}\ }\textbf {\bibinfo {volume} {609}},\ \bibinfo {pages} {131}
  (\bibinfo {year} {1996})}\BibitemShut {NoStop}%
\bibitem [{\citenamefont {Meng}\ \emph
  {et~al.}(2006{\natexlab{b}})\citenamefont {Meng}, \citenamefont {Peng},
  \citenamefont {Zhang},\ and\ \citenamefont {Zhou}}]{Meng2006_PRC73-037303}%
  \BibitemOpen
  \bibfield  {author} {\bibinfo {author} {\bibfnamefont {J.}~\bibnamefont
  {Meng}}, \bibinfo {author} {\bibfnamefont {J.}~\bibnamefont {Peng}}, \bibinfo
  {author} {\bibfnamefont {S.~Q.}\ \bibnamefont {Zhang}}, \ and\ \bibinfo
  {author} {\bibfnamefont {S.-G.}\ \bibnamefont {Zhou}},\ }\href {\doibase
  10.1103/PhysRevC.73.037303} {\bibfield  {journal} {\bibinfo  {journal} {Phys.
  Rev. C}\ }\textbf {\bibinfo {volume} {73}},\ \bibinfo {pages} {037303}
  (\bibinfo {year} {2006}{\natexlab{b}})}\BibitemShut {NoStop}%
\bibitem [{\citenamefont {Gambhir}\ \emph {et~al.}(1990)\citenamefont
  {Gambhir}, \citenamefont {Ring},\ and\ \citenamefont
  {Thimet}}]{Gambhir1990_AoP198-132}%
  \BibitemOpen
  \bibfield  {author} {\bibinfo {author} {\bibfnamefont {Y.~K.}\ \bibnamefont
  {Gambhir}}, \bibinfo {author} {\bibfnamefont {P.}~\bibnamefont {Ring}}, \
  and\ \bibinfo {author} {\bibfnamefont {A.}~\bibnamefont {Thimet}},\ }\href
  {\doibase 10.1016/0003-4916(90)90330-Q} {\bibfield  {journal} {\bibinfo
  {journal} {Ann. Phys.}\ }\textbf {\bibinfo {volume} {198}},\ \bibinfo {pages}
  {132} (\bibinfo {year} {1990})}\BibitemShut {NoStop}%
\bibitem [{\citenamefont {Ring}\ \emph {et~al.}(1997)\citenamefont {Ring},
  \citenamefont {Gambhir},\ and\ \citenamefont
  {Lalazissis}}]{Ring1997_CPC105-77}%
  \BibitemOpen
  \bibfield  {author} {\bibinfo {author} {\bibfnamefont {P.}~\bibnamefont
  {Ring}}, \bibinfo {author} {\bibfnamefont {Y.~K.}\ \bibnamefont {Gambhir}}, \
  and\ \bibinfo {author} {\bibfnamefont {G.~A.}\ \bibnamefont {Lalazissis}},\
  }\href {\doibase 10.1016/S0010-4655(97)00022-2} {\bibfield  {journal}
  {\bibinfo  {journal} {Comput. Phys. Commun.}\ }\textbf {\bibinfo {volume}
  {105}},\ \bibinfo {pages} {77} (\bibinfo {year} {1997})}\BibitemShut
  {NoStop}%
\bibitem [{\citenamefont {Song}\ \emph {et~al.}(2010)\citenamefont {Song},
  \citenamefont {Yao}, \citenamefont {L\"{u}},\ and\ \citenamefont
  {Meng}}]{Song2010_IJMPE19-2538}%
  \BibitemOpen
  \bibfield  {author} {\bibinfo {author} {\bibfnamefont {C.~Y.}\ \bibnamefont
  {Song}}, \bibinfo {author} {\bibfnamefont {J.~M.}\ \bibnamefont {Yao}},
  \bibinfo {author} {\bibfnamefont {H.~F.}\ \bibnamefont {L\"{u}}}, \ and\
  \bibinfo {author} {\bibfnamefont {J.}~\bibnamefont {Meng}},\ }\href {\doibase
  10.1142/S0218301310017058} {\bibfield  {journal} {\bibinfo  {journal} {Int.
  J. Mod. Phys. E}\ }\textbf {\bibinfo {volume} {19}},\ \bibinfo {pages} {2538}
  (\bibinfo {year} {2010})}\BibitemShut {NoStop}%
\bibitem [{\citenamefont {Long}\ \emph {et~al.}(2004)\citenamefont {Long},
  \citenamefont {Meng}, \citenamefont {Giai},\ and\ \citenamefont
  {Zhou}}]{Long2004_PRC69-034319}%
  \BibitemOpen
  \bibfield  {author} {\bibinfo {author} {\bibfnamefont {W.}~\bibnamefont
  {Long}}, \bibinfo {author} {\bibfnamefont {J.}~\bibnamefont {Meng}}, \bibinfo
  {author} {\bibfnamefont {N.~V.}\ \bibnamefont {Giai}}, \ and\ \bibinfo
  {author} {\bibfnamefont {S.-G.}\ \bibnamefont {Zhou}},\ }\href {\doibase
  10.1103/PhysRevC.69.034319} {\bibfield  {journal} {\bibinfo  {journal} {Phys.
  Rev. C}\ }\textbf {\bibinfo {volume} {69}},\ \bibinfo {pages} {034319}
  (\bibinfo {year} {2004})}\BibitemShut {NoStop}%
\bibitem [{\citenamefont {Sharma}\ \emph {et~al.}(1993)\citenamefont {Sharma},
  \citenamefont {Nagarajan},\ and\ \citenamefont
  {Ring}}]{Sharma1993_PLB312-377}%
  \BibitemOpen
  \bibfield  {author} {\bibinfo {author} {\bibfnamefont {M.~M.}\ \bibnamefont
  {Sharma}}, \bibinfo {author} {\bibfnamefont {M.~A.}\ \bibnamefont
  {Nagarajan}}, \ and\ \bibinfo {author} {\bibfnamefont {P.}~\bibnamefont
  {Ring}},\ }\href
  {http://www.sciencedirect.com/science/article/B6TVN-470G13S-69/2/b3e2b625f3a9be50fc271b3986c87fda}
  {\bibfield  {journal} {\bibinfo  {journal} {Phys. Lett. B}\ }\textbf
  {\bibinfo {volume} {312}},\ \bibinfo {pages} {377} (\bibinfo {year}
  {1993})}\BibitemShut {NoStop}%
\bibitem [{\citenamefont {Moeller}\ and\ \citenamefont
  {Nix}(1992)}]{Moeller1992_NPA536-20}%
  \BibitemOpen
  \bibfield  {author} {\bibinfo {author} {\bibfnamefont {P.}~\bibnamefont
  {Moeller}}\ and\ \bibinfo {author} {\bibfnamefont {J.~R.}\ \bibnamefont
  {Nix}},\ }\href {\doibase 10.1016/0375-9474(92)90244-E} {\bibfield  {journal}
  {\bibinfo  {journal} {Nucl. Phys. A}\ }\textbf {\bibinfo {volume} {536}},\
  \bibinfo {pages} {20} (\bibinfo {year} {1992})}\BibitemShut {NoStop}%
\bibitem [{\citenamefont {Karatzikos}\ \emph {et~al.}(2010)\citenamefont
  {Karatzikos}, \citenamefont {Afanasjev}, \citenamefont {Lalazissis},\ and\
  \citenamefont {Ring}}]{Karatzikos2010_PLB689-72}%
  \BibitemOpen
  \bibfield  {author} {\bibinfo {author} {\bibfnamefont {S.}~\bibnamefont
  {Karatzikos}}, \bibinfo {author} {\bibfnamefont {A.}~\bibnamefont
  {Afanasjev}}, \bibinfo {author} {\bibfnamefont {G.}~\bibnamefont
  {Lalazissis}}, \ and\ \bibinfo {author} {\bibfnamefont {P.}~\bibnamefont
  {Ring}},\ }\href {\doibase 10.1016/j.physletb.2010.04.045} {\bibfield
  {journal} {\bibinfo  {journal} {Phys. Lett. B}\ }\textbf {\bibinfo {volume}
  {689}},\ \bibinfo {pages} {72} (\bibinfo {year} {2010})}\BibitemShut
  {NoStop}%
\bibitem [{\citenamefont {Sharma}\ \emph {et~al.}(1999)\citenamefont {Sharma},
  \citenamefont {Mythili},\ and\ \citenamefont
  {Farhan}}]{Sharma1999_PRC59-1379}%
  \BibitemOpen
  \bibfield  {author} {\bibinfo {author} {\bibfnamefont {M.~M.}\ \bibnamefont
  {Sharma}}, \bibinfo {author} {\bibfnamefont {S.}~\bibnamefont {Mythili}}, \
  and\ \bibinfo {author} {\bibfnamefont {A.~R.}\ \bibnamefont {Farhan}},\
  }\href {\doibase 10.1103/PhysRevC.59.1379} {\bibfield  {journal} {\bibinfo
  {journal} {Phys. Rev. C}\ }\textbf {\bibinfo {volume} {59}},\ \bibinfo
  {pages} {1379} (\bibinfo {year} {1999})}\BibitemShut {NoStop}%
\bibitem [{\citenamefont {Lalazissis}\ \emph {et~al.}(2004)\citenamefont
  {Lalazissis}, \citenamefont {Vretenar},\ and\ \citenamefont
  {Ring}}]{Lalazissis2004_EPJA22-37}%
  \BibitemOpen
  \bibfield  {author} {\bibinfo {author} {\bibfnamefont {G.~A.}\ \bibnamefont
  {Lalazissis}}, \bibinfo {author} {\bibfnamefont {D.}~\bibnamefont
  {Vretenar}}, \ and\ \bibinfo {author} {\bibfnamefont {P.}~\bibnamefont
  {Ring}},\ }\href {\doibase 10.1140/epja/i2003-10227-7} {\bibfield  {journal}
  {\bibinfo  {journal} {Eur. Phys. J. A}\ }\textbf {\bibinfo {volume} {22}},\
  \bibinfo {pages} {37} (\bibinfo {year} {2004})}\BibitemShut {NoStop}%
\bibitem [{\citenamefont {Gangopadhyay}\ and\ \citenamefont
  {Roy}(2005)}]{Gangopadhyay2005_JPG31-1111}%
  \BibitemOpen
  \bibfield  {author} {\bibinfo {author} {\bibfnamefont {G.}~\bibnamefont
  {Gangopadhyay}}\ and\ \bibinfo {author} {\bibfnamefont {S.}~\bibnamefont
  {Roy}},\ }\href {\doibase 10.1088/0954-3899/31/9/012} {\bibfield  {journal}
  {\bibinfo  {journal} {J. Phys. G: Nucl. Phys.}\ }\textbf {\bibinfo {volume}
  {31}},\ \bibinfo {pages} {1111} (\bibinfo {year} {2005})}\BibitemShut
  {NoStop}%
\bibitem [{\citenamefont {Sagawa}\ \emph {et~al.}(2004)\citenamefont {Sagawa},
  \citenamefont {Zhou}, \citenamefont {Zhang},\ and\ \citenamefont
  {Suzuki}}]{Sagawa2004_PRC70-054316}%
  \BibitemOpen
  \bibfield  {author} {\bibinfo {author} {\bibfnamefont {H.}~\bibnamefont
  {Sagawa}}, \bibinfo {author} {\bibfnamefont {X.~R.}\ \bibnamefont {Zhou}},
  \bibinfo {author} {\bibfnamefont {X.~Z.}\ \bibnamefont {Zhang}}, \ and\
  \bibinfo {author} {\bibfnamefont {T.}~\bibnamefont {Suzuki}},\ }\href
  {\doibase 10.1103/PhysRevC.70.054316} {\bibfield  {journal} {\bibinfo
  {journal} {Phys. Rev. C}\ }\textbf {\bibinfo {volume} {70}},\ \bibinfo
  {pages} {054316} (\bibinfo {year} {2004})}\BibitemShut {NoStop}%
\bibitem [{\citenamefont {Zhang}\ \emph {et~al.}(2008)\citenamefont {Zhang},
  \citenamefont {Sagawa}, \citenamefont {Yoshino}, \citenamefont {Hagino},\
  and\ \citenamefont {Meng}}]{Zhang2008_PTP120-129}%
  \BibitemOpen
  \bibfield  {author} {\bibinfo {author} {\bibfnamefont {Y.}~\bibnamefont
  {Zhang}}, \bibinfo {author} {\bibfnamefont {H.}~\bibnamefont {Sagawa}},
  \bibinfo {author} {\bibfnamefont {D.}~\bibnamefont {Yoshino}}, \bibinfo
  {author} {\bibfnamefont {K.}~\bibnamefont {Hagino}}, \ and\ \bibinfo {author}
  {\bibfnamefont {J.}~\bibnamefont {Meng}},\ }\href {\doibase
  10.1143/PTP.120.129} {\bibfield  {journal} {\bibinfo  {journal} {Prog. Theor.
  Phys.}\ }\textbf {\bibinfo {volume} {120}},\ \bibinfo {pages} {129} (\bibinfo
  {year} {2008})}\BibitemShut {NoStop}%
\bibitem [{\citenamefont {Specht}\ \emph {et~al.}(1971)\citenamefont {Specht},
  \citenamefont {Schweimer}, \citenamefont {Rebel}, \citenamefont {Schatz},
  \citenamefont {Lohken},\ and\ \citenamefont {Hauser}}]{Specht1971_NPA171-65}%
  \BibitemOpen
  \bibfield  {author} {\bibinfo {author} {\bibfnamefont {J.}~\bibnamefont
  {Specht}}, \bibinfo {author} {\bibfnamefont {G.~W.}\ \bibnamefont
  {Schweimer}}, \bibinfo {author} {\bibfnamefont {H.}~\bibnamefont {Rebel}},
  \bibinfo {author} {\bibfnamefont {G.}~\bibnamefont {Schatz}}, \bibinfo
  {author} {\bibfnamefont {R.}~\bibnamefont {Lohken}}, \ and\ \bibinfo {author}
  {\bibfnamefont {G.}~\bibnamefont {Hauser}},\ }\href {\doibase
  10.1016/0375-9474(71)90363-0} {\bibfield  {journal} {\bibinfo  {journal}
  {Nucl. Phys. A}\ }\textbf {\bibinfo {volume} {171}},\ \bibinfo {pages} {65}
  (\bibinfo {year} {1971})}\BibitemShut {NoStop}%
\bibitem [{\citenamefont {Yasue}\ \emph {et~al.}(1983)\citenamefont {Yasue},
  \citenamefont {Tanabe}, \citenamefont {Soga}, \citenamefont {Kokame},
  \citenamefont {Shimokoshi}, \citenamefont {Kasagi}, \citenamefont {Toba},
  \citenamefont {Kadota}, \citenamefont {Ohsawa},\ and\ \citenamefont
  {Furuno}}]{Yasue1983_NPA394-29}%
  \BibitemOpen
  \bibfield  {author} {\bibinfo {author} {\bibfnamefont {M.}~\bibnamefont
  {Yasue}}, \bibinfo {author} {\bibfnamefont {T.}~\bibnamefont {Tanabe}},
  \bibinfo {author} {\bibfnamefont {F.}~\bibnamefont {Soga}}, \bibinfo {author}
  {\bibfnamefont {J.}~\bibnamefont {Kokame}}, \bibinfo {author} {\bibfnamefont
  {F.}~\bibnamefont {Shimokoshi}}, \bibinfo {author} {\bibfnamefont
  {J.}~\bibnamefont {Kasagi}}, \bibinfo {author} {\bibfnamefont
  {Y.}~\bibnamefont {Toba}}, \bibinfo {author} {\bibfnamefont {Y.}~\bibnamefont
  {Kadota}}, \bibinfo {author} {\bibfnamefont {T.}~\bibnamefont {Ohsawa}}, \
  and\ \bibinfo {author} {\bibfnamefont {K.}~\bibnamefont {Furuno}},\ }\href
  {\doibase 10.1016/0375-9474(83)90159-8} {\bibfield  {journal} {\bibinfo
  {journal} {Nucl. Phys. A}\ }\textbf {\bibinfo {volume} {394}},\ \bibinfo
  {pages} {29} (\bibinfo {year} {1983})}\BibitemShut {NoStop}%
\bibitem [{\citenamefont {Simmonds}\ \emph {et~al.}(1988)\citenamefont
  {Simmonds}, \citenamefont {Pearce}, \citenamefont {Hayes}, \citenamefont
  {Clarice}, \citenamefont {Griffiths}, \citenamefont {Mannion},\ and\
  \citenamefont {Ogilvie}}]{Simmonds1988_NPA482-653}%
  \BibitemOpen
  \bibfield  {author} {\bibinfo {author} {\bibfnamefont {P.~J.}\ \bibnamefont
  {Simmonds}}, \bibinfo {author} {\bibfnamefont {K.~I.}\ \bibnamefont
  {Pearce}}, \bibinfo {author} {\bibfnamefont {P.~R.}\ \bibnamefont {Hayes}},
  \bibinfo {author} {\bibfnamefont {N.~M.}\ \bibnamefont {Clarice}}, \bibinfo
  {author} {\bibfnamefont {R.~J.}\ \bibnamefont {Griffiths}}, \bibinfo {author}
  {\bibfnamefont {M.~C.}\ \bibnamefont {Mannion}}, \ and\ \bibinfo {author}
  {\bibfnamefont {C.~A.}\ \bibnamefont {Ogilvie}},\ }\href {\doibase
  10.1016/0375-9474(88)90174-1} {\bibfield  {journal} {\bibinfo  {journal}
  {Nucl. Phys. A}\ }\textbf {\bibinfo {volume} {482}},\ \bibinfo {pages} {653}
  (\bibinfo {year} {1988})}\BibitemShut {NoStop}%
\bibitem [{\citenamefont {Buervenich}\ \emph {et~al.}(2008)\citenamefont
  {Buervenich}, \citenamefont {Guo}, \citenamefont {Reinhard},\ and\
  \citenamefont {Greiner}}]{Buervenich2008_JPG35-025103}%
  \BibitemOpen
  \bibfield  {author} {\bibinfo {author} {\bibfnamefont {T.~J.}\ \bibnamefont
  {Buervenich}}, \bibinfo {author} {\bibfnamefont {L.}~\bibnamefont {Guo}},
  \bibinfo {author} {\bibfnamefont {P.-G.}\ \bibnamefont {Reinhard}}, \ and\
  \bibinfo {author} {\bibfnamefont {W.}~\bibnamefont {Greiner}},\ }\href
  {\doibase 10.1088/0954-3899/35/2/025103} {\bibfield  {journal} {\bibinfo
  {journal} {J. Phys. G: Nucl. Part. Phys.}\ }\textbf {\bibinfo {volume}
  {35}},\ \bibinfo {pages} {025103} (\bibinfo {year} {2008})}\BibitemShut
  {NoStop}%
\bibitem [{\citenamefont {Bender}\ \emph {et~al.}(2003)\citenamefont {Bender},
  \citenamefont {Heenen},\ and\ \citenamefont
  {Reinhard}}]{Bender2003_RMP75-121}%
  \BibitemOpen
  \bibfield  {author} {\bibinfo {author} {\bibfnamefont {M.}~\bibnamefont
  {Bender}}, \bibinfo {author} {\bibfnamefont {P.-H.}\ \bibnamefont {Heenen}},
  \ and\ \bibinfo {author} {\bibfnamefont {P.-G.}\ \bibnamefont {Reinhard}},\
  }\href {\doibase 10.1103/RevModPhys.75.121} {\bibfield  {journal} {\bibinfo
  {journal} {Rev. Mod. Phys.}\ }\textbf {\bibinfo {volume} {75}},\ \bibinfo
  {pages} {121} (\bibinfo {year} {2003})}\BibitemShut {NoStop}%
\bibitem [{\citenamefont {Yao}\ \emph {et~al.}(2010)\citenamefont {Yao},
  \citenamefont {Meng}, \citenamefont {Ring},\ and\ \citenamefont
  {Vretenar}}]{Yao2010_PRC81-044311}%
  \BibitemOpen
  \bibfield  {author} {\bibinfo {author} {\bibfnamefont {J.~M.}\ \bibnamefont
  {Yao}}, \bibinfo {author} {\bibfnamefont {J.}~\bibnamefont {Meng}}, \bibinfo
  {author} {\bibfnamefont {P.}~\bibnamefont {Ring}}, \ and\ \bibinfo {author}
  {\bibfnamefont {D.}~\bibnamefont {Vretenar}},\ }\href {\doibase
  10.1103/PhysRevC.81.044311} {\bibfield  {journal} {\bibinfo  {journal} {Phys.
  Rev. C}\ }\textbf {\bibinfo {volume} {81}},\ \bibinfo {pages} {044311}
  (\bibinfo {year} {2010})}\BibitemShut {NoStop}%
\bibitem [{\citenamefont {Yao}\ \emph {et~al.}(2011{\natexlab{a}})\citenamefont
  {Yao}, \citenamefont {Mei}, \citenamefont {Chen}, \citenamefont {Meng},
  \citenamefont {Ring},\ and\ \citenamefont {Vretenar}}]{Yao2011_PRC83-014308}%
  \BibitemOpen
  \bibfield  {author} {\bibinfo {author} {\bibfnamefont {J.~M.}\ \bibnamefont
  {Yao}}, \bibinfo {author} {\bibfnamefont {H.}~\bibnamefont {Mei}}, \bibinfo
  {author} {\bibfnamefont {H.}~\bibnamefont {Chen}}, \bibinfo {author}
  {\bibfnamefont {J.}~\bibnamefont {Meng}}, \bibinfo {author} {\bibfnamefont
  {P.}~\bibnamefont {Ring}}, \ and\ \bibinfo {author} {\bibfnamefont
  {D.}~\bibnamefont {Vretenar}},\ }\href {\doibase 10.1103/PhysRevC.83.014308}
  {\bibfield  {journal} {\bibinfo  {journal} {Phys. Rev. C}\ }\textbf {\bibinfo
  {volume} {83}},\ \bibinfo {pages} {014308} (\bibinfo {year}
  {2011}{\natexlab{a}})}\BibitemShut {NoStop}%
\bibitem [{\citenamefont {Tanaka}\ \emph {et~al.}(2010)\citenamefont {Tanaka},
  \citenamefont {Yamaguchi}, \citenamefont {Suzuki}, \citenamefont {Ohtsubo},
  \citenamefont {Fukuda}, \citenamefont {Nishimura}, \citenamefont {Takechi},
  \citenamefont {Ogata}, \citenamefont {Ozawa}, \citenamefont {Izumikawa},
  \citenamefont {Aiba}, \citenamefont {Aoi}, \citenamefont {Baba},
  \citenamefont {Hashizume}, \citenamefont {Inafuku}, \citenamefont {Iwasa},
  \citenamefont {Kobayashi}, \citenamefont {Komuro}, \citenamefont {Kondo},
  \citenamefont {Kubo}, \citenamefont {Kurokawa}, \citenamefont {Matsuyama},
  \citenamefont {Michimasa}, \citenamefont {Motobayashi}, \citenamefont
  {Nakabayashi}, \citenamefont {Nakajima}, \citenamefont {Nakamura},
  \citenamefont {Sakurai}, \citenamefont {Shinoda}, \citenamefont {Shinohara},
  \citenamefont {Suzuki}, \citenamefont {Takeshita}, \citenamefont {Takeuchi},
  \citenamefont {Togano}, \citenamefont {Yamada}, \citenamefont {Yasuno},\ and\
  \citenamefont {Yoshitake}}]{Tanaka2010_PRL104-062701}%
  \BibitemOpen
  \bibfield  {author} {\bibinfo {author} {\bibfnamefont {K.}~\bibnamefont
  {Tanaka}}, \bibinfo {author} {\bibfnamefont {T.}~\bibnamefont {Yamaguchi}},
  \bibinfo {author} {\bibfnamefont {T.}~\bibnamefont {Suzuki}}, \bibinfo
  {author} {\bibfnamefont {T.}~\bibnamefont {Ohtsubo}}, \bibinfo {author}
  {\bibfnamefont {M.}~\bibnamefont {Fukuda}}, \bibinfo {author} {\bibfnamefont
  {D.}~\bibnamefont {Nishimura}}, \bibinfo {author} {\bibfnamefont
  {M.}~\bibnamefont {Takechi}}, \bibinfo {author} {\bibfnamefont
  {K.}~\bibnamefont {Ogata}}, \bibinfo {author} {\bibfnamefont
  {A.}~\bibnamefont {Ozawa}}, \bibinfo {author} {\bibfnamefont
  {T.}~\bibnamefont {Izumikawa}}, \bibinfo {author} {\bibfnamefont
  {T.}~\bibnamefont {Aiba}}, \bibinfo {author} {\bibfnamefont {N.}~\bibnamefont
  {Aoi}}, \bibinfo {author} {\bibfnamefont {H.}~\bibnamefont {Baba}}, \bibinfo
  {author} {\bibfnamefont {Y.}~\bibnamefont {Hashizume}}, \bibinfo {author}
  {\bibfnamefont {K.}~\bibnamefont {Inafuku}}, \bibinfo {author} {\bibfnamefont
  {N.}~\bibnamefont {Iwasa}}, \bibinfo {author} {\bibfnamefont
  {K.}~\bibnamefont {Kobayashi}}, \bibinfo {author} {\bibfnamefont
  {M.}~\bibnamefont {Komuro}}, \bibinfo {author} {\bibfnamefont
  {Y.}~\bibnamefont {Kondo}}, \bibinfo {author} {\bibfnamefont
  {T.}~\bibnamefont {Kubo}}, \bibinfo {author} {\bibfnamefont {M.}~\bibnamefont
  {Kurokawa}}, \bibinfo {author} {\bibfnamefont {T.}~\bibnamefont {Matsuyama}},
  \bibinfo {author} {\bibfnamefont {S.}~\bibnamefont {Michimasa}}, \bibinfo
  {author} {\bibfnamefont {T.}~\bibnamefont {Motobayashi}}, \bibinfo {author}
  {\bibfnamefont {T.}~\bibnamefont {Nakabayashi}}, \bibinfo {author}
  {\bibfnamefont {S.}~\bibnamefont {Nakajima}}, \bibinfo {author}
  {\bibfnamefont {T.}~\bibnamefont {Nakamura}}, \bibinfo {author}
  {\bibfnamefont {H.}~\bibnamefont {Sakurai}}, \bibinfo {author} {\bibfnamefont
  {R.}~\bibnamefont {Shinoda}}, \bibinfo {author} {\bibfnamefont
  {M.}~\bibnamefont {Shinohara}}, \bibinfo {author} {\bibfnamefont
  {H.}~\bibnamefont {Suzuki}}, \bibinfo {author} {\bibfnamefont
  {E.}~\bibnamefont {Takeshita}}, \bibinfo {author} {\bibfnamefont
  {S.}~\bibnamefont {Takeuchi}}, \bibinfo {author} {\bibfnamefont
  {Y.}~\bibnamefont {Togano}}, \bibinfo {author} {\bibfnamefont
  {K.}~\bibnamefont {Yamada}}, \bibinfo {author} {\bibfnamefont
  {T.}~\bibnamefont {Yasuno}}, \ and\ \bibinfo {author} {\bibfnamefont
  {M.}~\bibnamefont {Yoshitake}},\ }\href {\doibase
  10.1103/PhysRevLett.104.062701} {\bibfield  {journal} {\bibinfo  {journal}
  {Phys. Rev. Lett.}\ }\textbf {\bibinfo {volume} {104}},\ \bibinfo {pages}
  {062701} (\bibinfo {year} {2010})}\BibitemShut {NoStop}%
\bibitem [{\citenamefont {Zhou}\ \emph {et~al.}(2000)\citenamefont {Zhou},
  \citenamefont {Meng}, \citenamefont {Yamaji},\ and\ \citenamefont
  {Yang}}]{Zhou2000_CPL17-717}%
  \BibitemOpen
  \bibfield  {author} {\bibinfo {author} {\bibfnamefont {S.-G.}\ \bibnamefont
  {Zhou}}, \bibinfo {author} {\bibfnamefont {J.}~\bibnamefont {Meng}}, \bibinfo
  {author} {\bibfnamefont {S.}~\bibnamefont {Yamaji}}, \ and\ \bibinfo {author}
  {\bibfnamefont {S.-C.}\ \bibnamefont {Yang}},\ }\href {\doibase
  10.1088/0256-307X/17/10/006} {\bibfield  {journal} {\bibinfo  {journal}
  {Chin. Phys. Lett.}\ }\textbf {\bibinfo {volume} {17}},\ \bibinfo {pages}
  {717} (\bibinfo {year} {2000})}\BibitemShut {NoStop}%
\bibitem [{\citenamefont {Stoitsov}\ \emph {et~al.}(2000)\citenamefont
  {Stoitsov}, \citenamefont {Dobaczewski}, \citenamefont {Ring},\ and\
  \citenamefont {Pittel}}]{Stoitsov2000_PRC61-034311}%
  \BibitemOpen
  \bibfield  {author} {\bibinfo {author} {\bibfnamefont {M.~V.}\ \bibnamefont
  {Stoitsov}}, \bibinfo {author} {\bibfnamefont {J.}~\bibnamefont
  {Dobaczewski}}, \bibinfo {author} {\bibfnamefont {P.}~\bibnamefont {Ring}}, \
  and\ \bibinfo {author} {\bibfnamefont {S.}~\bibnamefont {Pittel}},\ }\href
  {\doibase 10.1103/PhysRevC.61.034311} {\bibfield  {journal} {\bibinfo
  {journal} {Phys. Rev. C}\ }\textbf {\bibinfo {volume} {61}},\ \bibinfo
  {pages} {034311} (\bibinfo {year} {2000})}\BibitemShut {NoStop}%
\bibitem [{\citenamefont {Zhou}\ \emph {et~al.}(2003)\citenamefont {Zhou},
  \citenamefont {Meng},\ and\ \citenamefont {Ring}}]{Zhou2003_PRC68-034323}%
  \BibitemOpen
  \bibfield  {author} {\bibinfo {author} {\bibfnamefont {S.-G.}\ \bibnamefont
  {Zhou}}, \bibinfo {author} {\bibfnamefont {J.}~\bibnamefont {Meng}}, \ and\
  \bibinfo {author} {\bibfnamefont {P.}~\bibnamefont {Ring}},\ }\href {\doibase
  10.1103/PhysRevC.68.034323} {\bibfield  {journal} {\bibinfo  {journal} {Phys.
  Rev. C}\ }\textbf {\bibinfo {volume} {68}},\ \bibinfo {pages} {034323}
  (\bibinfo {year} {2003})}\BibitemShut {NoStop}%
\bibitem [{\citenamefont {Zhou}\ \emph {et~al.}(2010)\citenamefont {Zhou},
  \citenamefont {Meng}, \citenamefont {Ring},\ and\ \citenamefont
  {Zhao}}]{Zhou2010_PRC82-011301R}%
  \BibitemOpen
  \bibfield  {author} {\bibinfo {author} {\bibfnamefont {S.-G.}\ \bibnamefont
  {Zhou}}, \bibinfo {author} {\bibfnamefont {J.}~\bibnamefont {Meng}}, \bibinfo
  {author} {\bibfnamefont {P.}~\bibnamefont {Ring}}, \ and\ \bibinfo {author}
  {\bibfnamefont {E.-G.}\ \bibnamefont {Zhao}},\ }\href {\doibase
  10.1103/PhysRevC.82.011301} {\bibfield  {journal} {\bibinfo  {journal} {Phys.
  Rev. C}\ }\textbf {\bibinfo {volume} {82}},\ \bibinfo {pages} {011301R}
  (\bibinfo {year} {2010})}\BibitemShut {NoStop}%
\bibitem [{\citenamefont {Ring}\ and\ \citenamefont {Schuck}(1980)}]{Ring1980}%
  \BibitemOpen
  \bibfield  {author} {\bibinfo {author} {\bibfnamefont {P.}~\bibnamefont
  {Ring}}\ and\ \bibinfo {author} {\bibfnamefont {P.}~\bibnamefont {Schuck}},\
  }\href@noop {} {\emph {\bibinfo {title} {The Nuclear Many-Body Problem}}}\
  (\bibinfo  {publisher} {Springer},\ \bibinfo {year} {1980})\BibitemShut
  {NoStop}%
\bibitem [{\citenamefont {Yao}\ \emph {et~al.}(2011{\natexlab{b}})\citenamefont
  {Yao}, \citenamefont {Li}, \citenamefont {Hagino}, \citenamefont {Thi~Win},
  \citenamefont {Zhang},\ and\ \citenamefont {Meng}}]{Yao2011_1104.3200}%
  \BibitemOpen
  \bibfield  {author} {\bibinfo {author} {\bibfnamefont {J.~M.}\ \bibnamefont
  {Yao}}, \bibinfo {author} {\bibfnamefont {Z.~P.}\ \bibnamefont {Li}},
  \bibinfo {author} {\bibfnamefont {K.}~\bibnamefont {Hagino}}, \bibinfo
  {author} {\bibfnamefont {M.}~\bibnamefont {Thi~Win}}, \bibinfo {author}
  {\bibfnamefont {Y.}~\bibnamefont {Zhang}}, \ and\ \bibinfo {author}
  {\bibfnamefont {J.}~\bibnamefont {Meng}},\ }\href
  {http://arxiv.org/abs/1104.3200v1} {\enquote {\bibinfo {title} {Impurity
  effect of lambda hyperon on collective excitations of atomic nuclei},}\
  }\bibinfo {howpublished} {arXiv:1104.3200v1 [nucl-th]} (\bibinfo {year}
  {2011}{\natexlab{b}})\BibitemShut {NoStop}%
\end{thebibliography}

%

\end{document}